\documentclass[]{interact}
\usepackage{xcolor}
\usepackage{graphicx}
\usepackage{tikz}
\usetikzlibrary{calc,shapes,backgrounds, arrows, positioning}
\usepackage[myheadings]{fullpage}
\usepackage{IEEEtrantools}
\usepackage{amsmath}
\usepackage{amssymb}
\usepackage[natbibapa]{apacite}
\usepackage{float}
\usepackage{todonotes}
\usepackage{enumitem}
\setlist{nosep}

\tikzstyle{item}=[rectangle,draw=black,fill=white,
                    font=\scriptsize\sffamily\bfseries,inner sep=6pt]
\tikzstyle{factor}=[ellipse,draw=black,fill=white,
                     font=\scriptsize\sffamily\bfseries,inner sep=-1pt]

\tikzstyle{arr}=[-latex,black,line width=1pt]
\tikzstyle{doublearr}=[latex-latex,black,line width=1pt]

\usepackage{epstopdf}
\usepackage[caption=false]{subfig}
\usepackage{float}
\usepackage{tikz}
\usetikzlibrary{calc,shapes,backgrounds, arrows, positioning, fit}
\usepackage[longnamesfirst,sort]{natbib}
\bibpunct[, ]{(}{)}{;}{a}{,}{,}
\renewcommand\bibfont{\fontsize{10}{12}\selectfont}
\usepackage{lineno}
\usepackage{setspace}

\theoremstyle{plain}

\theoremstyle{definition}

\theoremstyle{remark}

\begin{document}

\articletype{ORIGINAL ARTICLE}

\title{Optimal Instrument Selection using Bayesian Model Averaging for Model Implied Instrumental Variable Two Stage Least Squares Estimators}

  \author{Teague R. Henry \thanks{We gratefully acknowledge support from National Institute of Mental Health (1R21MH119572-01), the Population Science Training Program (T32 HD007168) and general support of the Carolina Population Center (P2C HD050924).} \\
  \affil{Department of Psychology and School of Data Science, University of Virginia}
  Zachary F. Fisher \\
  \affil{College of Human Development and Family Studies, Penn State}
  Kenneth A. Bollen \\
  \affil{Department of Psychology and Neuroscience and Department of Sociology, University of North Carolina at Chapel Hill}
  }

\maketitle

\begin{abstract}
Model-Implied Instrumental Variable Two-Stage Least Squares (MIIV-2SLS) is a limited information, equation-by-equation, non-iterative estimator for latent variable models. Associated with this estimator are equation specific tests of model misspecification. One issue with equation specific tests is that they lack specificity, in that they indicate that some instruments are problematic without revealing which specific ones. Instruments that are poor predictors of their target variables (“weak instruments”) is a second potential problem. We propose a novel extension to detect instrument specific tests of misspecification and weak instruments. We term this the Model-Implied Instrumental Variable Two-Stage Bayesian Model Averaging (MIIV-2SBMA) estimator. We evaluate the performance of MIIV-2SBMA against MIIV-2SLS in a simulation study and show that it has comparable performance in terms of parameter estimation. Additionally, our instrument specific overidentification tests developed within the MIIV-2SBMA framework show increased power to detect specific problematic and weak instruments. Finally, we demonstrate MIIV-2SBMA using an empirical example.
{}
\end{abstract}

\newpage

\section{Introduction}

Model-Implied Instrumental Variable Two-Stage Least Squares (MIIV-2SLS) is a limited-information estimator for structural equation models with latent variables \citep{Bollen1996}. Compared to systemwide estimators (e.g. maximum likelihood (ML)), MIIV-2SLS has been shown to better isolate bias resulting from model misspecification \citep{Bollen2007,Bollen2001}. This property results from the equation-by-equation estimation approach in MIIV-2SLS, in which the full structural equation model is broken down into individual equations, each of which can be estimated separately. This feature, combined with the computational simplicity and relative efficiency  of model implied instrumental variable (MIIV) estimation methods, makes it an attractive alternative or complement to full information ML, generalized least squares or weighted least squares approaches. 

Before proceeding with the development of a Bayesian MIIV-2SLS estimator some background information on the MIIV-2SLS estimator is in order. As with many structural equation modeling endeavors the first step in fitting a model using MIIV-2SLS is that of model specification. First, a structural model is specified corresponding to a theory one wishes to test. Next, following the model specification step \textit{a latent to observed variable transformation} is used to eliminate the presence of latent variables and derive a set of observed variable estimating equations. These observed variable estimating equations resemble simple linear regressions, however, the latent-to-observed transformation is likely to induce a correlation between the equation error and explanatory variables, compromising the consistency and asymptotic unbiasedness of the OLS estimator. Instrumental estimators, on the other hand, represent one possible solution for estimating the unknown parameters in the transformed estimating equations 

Although only recently gaining popularity in psychology, instrumental variable estimators have a long history in other fields, including economics and sociology. In a typical instrumental variable application, one might be concerned that the equation error correlates with one or more of the covariates in a regression model (which would lead to bias in the parameter estimates). In this typical situation a researcher would try to identify a variable from outside the model which correlates with the dubious covariates but is uncorrelated with the equation error, and use this instrument (or instruments) to estimate versions of the covariate that are not correlated with equation error. These versions of the covariates can be used in place of the original covariates, and if all assumptions hold, will result in unbiased parameter estimates. While the use of auxillary instruments is necessary in a single regression equation situation, in an SEM framework the overall model consists of many equations which in turn, if the model is correctly specified, completely describes the relations between all observed variables. MIIV takes advantage of this property of SEM models, and instead of using auxillary variables as instruments, draws instruments for each equation from within the model itself. \citet[][p. 114]{Bollen1996} describes two methods for determining the MIIVs for each equation in a given SEM and \citet{Bollen2004} and \citet{miivsem} created algorithms to automatically implement them.  Based on the model specification, these MIIVs fulfill key requirements of instrumental variables. Another difference between MIIVs and the typical use case of instrumental variables is that there are often far more MIIVs per equation than are strictly necessary for identification. In more traditional applications of instrumental variables, the researcher searches for instruments that are external to the model and good instruments are often difficult to find.

Once the MIIVs for each equation are determined, they are used to compute intermediate estimates of the endogenous predictors (via OLS) within the equation (stage 1 of Two Stage Least Squares). Those intermediate estimates are then used to estimate the associations between the endogenous predictors and the dependent variable with each equation. If the MIIVs are valid instruments, then this MIIV-2SLS estimator is consistent and asymptotically unbiased on an equation by equation basis \citep{Bollen1996}. Because MIIV-2SLS operates equation by equation, this limited information approach to estimation has a marked advantage over full information estimators such as ML, in that model misspecification that impacts one equation is less likely to impact the estimation of other equations \citep{Bollen2007,bollen_robustness_2018}. 

While the MIIV-2SLS approach has several advantages over maximum likelihood estimation when model misspecification is present, there are a number of open questions in the MIIV literature regarding the relationship between model misspecification diagnostics and instrument quality. One consideration is that if the structural model is misspecified, then some MIIVs will be invalid (as they were derived under the assumption that the specified model is the true model). Analysts can apply an overidentification test to each overidentified equation to assess evidence of misspecification \citep{Kirby2010}. Here, rejection of the null hypothesis suggests model misspecification, as the model specification led to the equation-specific MIIVs. Despite the utility of these overidentification tests, it is not always clear where in the model the misspecification originated as the equation failing the overidentification test may not be the equation where the misspecification occurred. Furthermore, it is also unclear which MIIVs are responsible for the failed test as the null hypothesis states all the overidentifying constraints hold. 

This issue, combined with the usual large number of implied instruments makes it difficult to evaluate the quality of a set of MIIVs, and therefore evaluate and localize misspecification in the larger model. In addition to invalid MIIVs, one must also be concerned with the problem of weak instruments. Such instruments have  associations or partial associations of near $0$ with the endogenous covariates. Some weak MIIVs could also be invalid instruments and the inclusion of weak and invalid instruments can have a large impact on the bias of parameter estimates \citep{Madigan1994,Magdalinos1985}. Therefore, it is important not only to identify invalid MIIVs, but to select the subset of MIIVs that are strongest for a given equation.

The MIIV-2SLS estimator would be greatly strengthened if we knew which \textit{specific} MIIVs in those equations are invalid or weak. The current manuscript's unique contribution to the SEM estimation literature is to provide a variant of MIIV-Bayesian Model Averaging for Model Implied Instrumental Variable Two Stage Least Squares Estimators2SLS that allows researchers to identify specific sources of model misspecification at the level of the instrument rather than the equation, while simultaneously accounting for weak instruments. This estimator, which we term MIIV Two Stage Bayesian Model Averaging (MIIV-2SBMA) modifies the procedure developed by  \citet{Lenkoski2014}, using Bayesian model averaging to combine estimates from all possible subsets of MIIVs. Using this approach, we propose Bayesian variants of Sargan's $\chi^2$ Test \citep{Sargan1958} for detecting invalid instruments at the level of the instrument itself, rather than the equation. We propose this test for both univariate (i.e. one endogenous variable predicting the outcome within an equation) and multivariate settings (i.e. more than one endogenous variable predicting the outcome within an equation).  Additionally, we demonstrate the use of inclusion probabilities to detect weak instruments. We conduct a series of Monte Carlo experiments to evaluate the performance of MIIV-2SBMA and our misspecification tests, and demonstrate that our approach shows increased power to detect model misspecification and weak instruments without a corresponding increase in the bias or variance of the model estimates, and allows us to identify specific sources of model misspecification. Finally, we present an empirical example demonstrating the use of MIIV-2SBMA for estimating a two factor CFA and determining which error covariances need to be included. 
    
The rest of this manuscript is structured as follows. We begin by providing a description of the instrumental variable approach to simple linear regression and how MIIV-2SLS relates. From there, we describe a common misspecification test, Sargan's Test \citep{Sargan1958}, and describe issues caused by weak instruments. This is followed by a description of Bayesian model averaging using local empirical Bayes $g$-priors \citep{Liang2008} for a single endogenous predictor and a multivariate g-prior \citep{Brown1998} for more than one endogenous predictor. We then describe Bayesian Model Averaged Sargan's Tests as well as our novel contribution, Instrument Specific Sargan's Test. We present the results from a Monte Carlo experiment, demonstrating the performance of Instrument Specific Sargan's Tests under a number of different conditions.  Finally, we illustrate the use of these instrument specific tests using an empirical example.
    
\section{A Brief Description of Instrumental Variable Regression and MIIV-2SLS}

A more technical description of MIIV-2SLS is in the Supplementary Materials. What follows is a description of simple instrumental variable regression and how MIIV-2SLS relates, with the intention being to provide an overview of the instrumental variable approach for the unfamiliar reader. Readers can find extended discussion of instrumental variable methods in most econometric texts \citep[e.g.][]{davidson_estimation_1993}. In addition, more extensive overviews of MIIV-2SLS are in \citet{bollen_model_2019} and \citet{bollen_introduction_2021}.

Consider a simple linear regression:
\begin{equation*}
    Y = \beta X + \boldsymbol{\varepsilon}
\end{equation*}
where $\boldsymbol{\varepsilon}$ is a random variable with mean of 0 and variance of $\sigma^2$ and $X$ is a variable selected out of a set of possible predictors. One important assumption of linear regression is that the predictor $x$ is exogenous, in that it is not causally determined by any other variable under consideration. The assumption of exogeneity implies that $Cov(X, \boldsymbol{\varepsilon}) = 0$ or that the covariate is uncorrelated with the error term. When this assumption is violated, the resulting estimator of $\beta$, $\hat{\beta}$, will be biased, which in turn will increase the risk of a spurious finding (or false negative). 

The instrumental variable approach  \citep{Sargan1958} seeks to identify additional variables $\mathbf{Z}$, called instruments, that are correlated with $X$, but not correlated with $\boldsymbol{\varepsilon}$. In a standard instrumental variable regression approach, these instruments are auxiliary to the analysis at hand, and are not used as predictors of the outcome. An instrumental variable method to estimating regression equations, Two Stage Least Squares (2SLS) first attempts to create a version of $X$, $\hat{X}$, that is asymptotically uncorrelated with $\boldsymbol{\varepsilon}$ by regressing $X$ on the instrument or instruments and extracting the predicted value of $X$ as $\hat{X}$. $\hat{X}$ is then a transformed version of $\mathbf{Z}$, that, if $\mathbf{Z}$ are proper instruments, are asymptotically uncorrelated with $\mathbf{\varepsilon}$. 

Two questions immediately come to mind with 2SLS:
\begin{enumerate}
    \item What if variables within $\mathbf{Z}$ are in fact asymptotically associated with $\boldsymbol{\varepsilon}$?
    \item What if variables within $\mathbf{Z}$ are not a strong predictors of $X$?
\end{enumerate}

In this basic regression setting, the answer to the first question is simple: If variables within $\mathbf{Z}$ are related to $\boldsymbol{\varepsilon}$ then those variables are invalid instrumental variables and should not be used. In this setting, one can remove those invalid instruments from the list of proposed instruments, however in the MIIV setting, this is a more complicated situation. 

The answer to the second question is similarly simple: If variables within $\mathbf{Z}$ are not strongly related to $X$ conditional on other variables in $\mathbf{Z}$, they are \textit{weak instruments}. Weak instruments, while still technically instrumental variables (as long as they are valid as per the first question), do not provide accurate predictions of $X$. Again, in the 2SLS regression context, if one has the choice between a strong or weak instrument, one should always choose to use the instrument with a stronger association with the predictor variables. 

To summarize the use of instrumental variables in this basic regression example: The predictor $X$ needs to be uncorrelated with the error term $\boldsymbol{\varepsilon}$, else bias will result in the Ordinary Least Squares (OLS) estimates of the regression coefficient. 2SLS uses instrumental variables $\mathbf{Z}$, variables that are correlated with $X$ but uncorrelated with $\boldsymbol{\varepsilon}$, to estimate a version of $X$, $\hat{X}$, that is asymptotically uncorrelated with the error term. If a given variable within $\mathbf{Z}$ is not a proper instrument, then its use as such will result in biased estimation, while if a variable with $\mathbf{Z}$ is only a weak predictor of $X$, the estimates of the regression coefficients will be more variable \citep{Bound1995}. 

Up till now, we have been dealing with a single equation system, which necessitates the selection of auxiliary instrumental variables. In a structural equation modelling framework however, we have multiple sets of equations that together describe the relations between all the variables in the system. Model Implied Instrumental Variable SEM (MIIVSEM; \cite{Bollen1996}) uses the structural information from the model to identify variables within the system that can act as instruments for other variables, rather than recruit additional auxiliary variables from outside of the system. We illustrate this approach by way of example.

\tikzstyle{item}=[rectangle,draw=black,fill=white,
                    font=\scriptsize\sffamily\bfseries,inner sep=6pt]
\tikzstyle{factor}=[ellipse,draw=black,fill=white,
                     font=\bfseries,inner sep=6pt]
\tikzstyle{caption}=[font=\scriptsize\bfseries]
\tikzstyle{arr}=[-latex,black,line width=.5pt, font=\tiny, anchor=center]
\tikzstyle{doublearr}=[latex-latex,black,line width=.5pt, font=\scriptsize]
\tikzstyle{labBox}=[rectangle, draw=black, fill=white,inner sep = .8pt]

\begin{figure}[H]
\centering
\begin{tikzpicture}[scale=.8, node distance=1cm, auto]
\node[factor](f1) at (-3.5, 0) {$\eta_1$} ;
\node[factor](f2) at (1.5, 0) {$\eta_2$} ;
\node[item](y1) at (-5,-2) {$Y_1$};
\node[item](y2) at (-3.5,-2) {$Y_2$};
\node[item](y3) at (-2,-2) {$Y_3$};
\node[item](y5) at (0,-2) {$Y_4$};
\node[item](y6) at (1.5,-2) {$Y_5$};
\node[item](y7) at (3,-2) {$Y_6$};
\draw[arr](f1) to node[labBox] {$1$} (y1.north) ;
\draw[arr](f1) to node[labBox] {$\lambda_2$} (y2.north) ;
\draw[arr](f1) to node[labBox] {$\lambda_3$} (y3.north);
\draw[arr](f2) to node[labBox] {$1$} (y5.north);
\draw[arr](f2) to node[labBox] {$\lambda_5$} (y6.north);
\draw[arr](f2) to node[labBox] {$\lambda_6$} (y7.north);
\draw[doublearr, bend left](f1.north) to node[anchor=center, above, font=\tiny] {} (f2.north);
;

\node(var1)[font=\small, inner sep = .25pt] at (-4.5, 1) {$\sigma^2_{\eta_1}$};
\node(var2)[font=\small, inner sep = .25pt] at (2.5, 1) {$\sigma^2_{\eta_2}$};
\draw[arr](var1) to (f1);
\draw[arr](var2) to (f2);

\end{tikzpicture}
\label{fig:2facpath}
\caption{Path diagram of a two factor confirmatory factor model.}
\end{figure}

Consider the path diagram in Fig. \ref{fig:2facpath}. The 6 equations that define this system are:
\begin{align*}
    Y_1 &= \eta_1 + \boldsymbol{\varepsilon}_1 \\
    Y_2 &= \lambda_2\eta_1 + \boldsymbol{\varepsilon}_2 \\
    Y_3 &= \lambda_3\eta_1 + \boldsymbol{\varepsilon}_3 \\
    Y_4 &= \eta_2 + \boldsymbol{\varepsilon}_4 \\
    Y_5 &= \lambda_5\eta_2 + \boldsymbol{\varepsilon}_5 \\
    Y_6 &= \lambda_6\eta_2 + \boldsymbol{\varepsilon}_6
\end{align*}
where, by definition, $Cov(\eta, \boldsymbol{\varepsilon}) = 0$ for any choice of $\eta$ and $\boldsymbol{\varepsilon}$. MIIV-2SLS first replaces the latent variables with their scaling indicators, for example, $Y_2 = \lambda_2\eta_1 + \boldsymbol{\varepsilon}_2 $ becomes $Y_2 = \lambda_2Y_1  +\boldsymbol{\varepsilon}_2 -\lambda_2\boldsymbol{\varepsilon}_1 \boldsymbol{\varepsilon}_2$. However, this transformation results in each equation violating the exogeneity assumption, in that the scaling indicators $Y_1$ and $Y_4$ will be related to the equation level error in each equation they appear in. As such, the scaling indicators cannot simply be used as replacements for the latent variables, else the resulting OLS estimator of $\lambda$ will be biased. This is when the \textit{model implied} aspect of model implied instrumental variables comes into play. 

Consider now the equation for $Y_2$ with the scaling indicator substituted in for $\eta_1$, $$Y_2 = \lambda_2Y_1 + \boldsymbol{\varepsilon}_2 - \lambda_2\boldsymbol{\varepsilon}_1.$$ To take account of $Y_1$ correlation with the composite error and to apply 2SLS,  we need to identify an instrumental variable that is correlated with $Y_1$, but uncorrelated with $\boldsymbol{\varepsilon}_1$ and $\boldsymbol{\varepsilon}_2$. This can, of course, be done using auxillary instruments from outside of the set of variables in the model, but the structure of the model described in Fig. \ref{fig:2facpath} provides a different option. Here, based on the proposed model, $Y_3$,$Y_4$, $Y_5$ and $Y_6$ are all correlated with $Y_1$, but uncorrelated with $\boldsymbol{\varepsilon}_2$. \textit{Thus, if the model is correctly specified, $Y_3$, $Y_4$, $Y_5$ and $Y_6$ are all instruments for the $Y_2$ equation.}

With the MIIVs for the previous equation so identified, the resulting 2SLS for $Y_1$  is the predicted value of $Y_1$ when $Y_1$ is regressed on 
 $Y_3, Y_4, Y_5, \text{ and } Y_6$. Using this estimate of $Y_1$, the OLS estimator of $\lambda_2$ is, if the model has been specified correctly, an asymptotically unbiased estimator of the population $\lambda_2$. This process is then repeated for each equation.

Generally speaking, the MIIV-2SLS estimator for structural equation models follows the logic outlined above: First, latent variables are replaced with their observed scaling indicators minus its error term. If this transformation were applied blindly, then the resulting set of regression equations would end up having biased estimates, so from the proposed model structure, instrumental variables for each equation are identified and the 2SLS estimation is carried out. More complex SEMs, such as those with higher-order factors \citep{bollen_note_2002} or categorical variables \citep{fisher_instrumental_2020} can also be estimated using MIIV-2SLS, though the procedure is more complex . The resulting estimates from the MIIV-2SLS estimator are asymptotically consistent, making MIIV-2SLS an attractive alternative to maximum likelihood estimation, as MIIV-2SLS is non-iterative, will not run into issues with model convergence, and has a number of other advantages \citep{bollen_model_2019}.

However, the identification of the model implied instruments depends on a difficult assumption to meet: That the model is correctly specified. If the model is correct, then the model implied instruments will be, by definition, valid instruments. If the model is misspecified, the model implied instruments for one or more equations in the system will be invalid instruments. This is not a disadvantage for MIIV-2SLS, rather it provides an opportunity to identify where in the model misspecification occurs, a topic we turn to now.

\subsection{Sargan's $\chi^2$ Test of Overidentification}

Recall the requirement that an instrument must not correlate with the equation error. We term variables that violate this requirement but are still inappropriately used as instruments, \textit{invalid instruments}. Importantly, invalid instruments in the context of MIIVs arise when the model is misspecified. Although the validity of the assumption cannot be evaluated directly, we can assess the appropriateness of an instrument set in the context of an overidentified equation (i.e., one where the number of instruments exceed the number of endogenous predictors\footnote{The use of the term \textit{identification} in a maximum likelihood estimated SEM vs. MIIV-SEM setting are referring to related considerations. In the traditional ML setting, a model is overidentified if there are multiple ways to estimate one or more parameters. In the MIIV-SEM setting, this consideration is applied on a equation by equation basis, wherein an equation is overidentified if there are multiple ways of estimating the parameter values.}) using overidentification tests such as Sargan's $\chi^2$ test \citep{Sargan1958}. In the context of latent variable models Kirby and Bollen (\citeyear{Kirby2010}) found Sargan's Test performed better than other overidentification tests and we use it here.
	
If all MIIVs in an overidentified equation are valid instruments, then each overidentified coefficient should lead to the same value in the population.  Even if this true, sampling fluctuations can lead to different values. Sargan's Test of overidentification determines whether these different solutions are within sampling error.  Researchers can estimate the test statistic as $nR^2$, where $n$ is the sample size and $R^2$ is the squared multiple correlation from the regression of the equation sample residuals on the instruments and the resulting statistic asymptotically follows a $\chi^2$ distribution. The degrees of freedom associated with this distribution are equal to the degree of overidentification for the equation (i.e., $p_{\mathbf{Z}_j} - p_j$, where $p_{\mathbf{Z}_j}$ denotes the number of MIIVs and $p_j$ is the number of regressors). Rejection of the null hypothesis suggests that one or more of the instruments for that equation correlates with the equation error, and as such is an invalid instrument. In the MIIV setting, these tests  indicate whether the model is misspecified. After all, the instruments for any given equation are \textit{a priori} determined by the model structure, so if an instrument is invalid, the model structure itself is invalid \citep{Kirby2010}.
  
However, as was mentioned previously, Sargan's Test lacks the ability to pinpoint sources of model misspecification beyond the set of MIIVs of a specific equation. The Sargan's Test assesses if \textit{at least one} instrument is invalid. Though this is a local (equation) test of overidentification, it does not reveal \textit{which} of the MIIVs are the source of the problem. This issue is compounded in the MIIV setting as for each equation there are normally many MIIVs, and an omnibus test of instrument validity would not provide specific enough information as to the source of model misspecification. A more useful test would be one that identifies a specific failed instrument. We will show how our MIIV-2SBMA approach can better isolate the problematic instruments.

\subsection{Weak Instruments}

Complementary to invalid instruments that correlate with the equation error are weak instruments which are weakly associated with the regressors that they are to predict.  The inclusion of weak, invalid instruments in the 2SLS estimator leads to inconsistent estimates \citep{Bound1995} as well as increasing the bias of the estimator \citep{Mariano1977,Magdalinos1985}. Therefore, weak instruments are to be avoided. While weak instruments are less of a threat to consistency than invalid instruments, they are still a concern particularly in the MIIV setting. Again, for any given equation, there are typically multiple MIIVs, and their number grows large when there are multiple indicators. It is not uncommon to see five or six MIIVs for each equation, while in the more traditional instrumental variable regression setting having more than one or two auxillary instruments is rare. Given this, and the fact that for any reasonably sized SEM there are multiple equations to consider, having an automated approach to identifying and accounting for weak instruments would improve the usability of the overall method.
    
\subsection{Methodological Goals}
With both Sargan's Test and weak instruments described, we have outlined the needed methodological advances:
\begin{enumerate}
\item{Analysts using MIIV-2SLS should be able to find the invalid MIIVs rather than just showing at least one MIIV is a problem.}
\item{They should also be able to identify and account for weak instruments, without needing to manually select MIIVs for each equation.}
\end{enumerate}

To make these advances, we now turn to Bayesian Model Averaging (BMA; \cite{Raftery1997}). We first describe BMA generally, and then describe how it has been used to perform instrument selection in the traditional instrumental variable regression setting. From there, we provide an extension to MIIVSEM that allows for the instrument level model misspecification identification, while simultaneously accounting for weak instruments.

\section{Bayesian Model Averaging}

Bayesian model averaging (BMA) is a powerful tool for quantifying  uncertainty in model specification \citep{Raftery1997} and typically leads to improved prediction and better parameter estimates than any single model \citep{Madigan1994}. At a high level, BMA seeks to estimate the posterior distribution of a parameter of interest $\theta$ in the following fashion
\begin{equation}
\mathbb{P}(\theta|D) = \sum^{K}_{k = 1} \mathbb{P}(\theta|M_k, D)\mathbb{P}(M_k|D) 
\end{equation}
where $D$ is the observed data, and $M_k$ is the $k$th model from some set of models of size $K$, $\mathcal{M}$. The form of any $M_k$ depends on the specific application. In a linear regression context, $\mathcal{M}$ usually consists of models evaluating every possible subset of predictors. 

In practice, the posterior probability for a specific model,  $M_k$, $\mathbb{P}(M_k|D)$ is calculated by first calculating a \textit{Bayes Factor}, which is the ratio of two models' likelihoods
\begin{equation}
BF[M_k:M_l] = \frac{\mathbb{P}(D|M_k)}{\mathbb{P}(D|M_l)} = \frac{\mathbb{P}(M_k|D)\mathbb{P}(M_k)}{\mathbb{P}(M_l|D)\mathbb{P}(M_l)}.
\end{equation}

When using Bayes Factors for model averaging purposes, a common comparison model allows $\mathcal{P}(M_k|D)$ to be calculated. One common choice is that of the \textit{null model}, $M_0$, the model that assumes no relation between variables. In the context of linear regression, the null model consists of solely the intercept.

Given a set of models $\{M_1, \dots, M_K\}$ the posterior probability of any model $M_k$ relative to the set of models given data $D$ is
\begin{equation}
\mathbb{P}(M_k|D) = \frac{\mathbb{P}(M_k)BF[M_k:M_0]}{\sum_{l = 1}^K \mathbb{P}(M_l)BF[M_l:M_0]}.
\end{equation}

If the prior probability of any model $\mathcal{P}(M_k)$ is equal to some constant (i.e., $\frac{1}{K}$), this simplifies to
\begin{equation}
\mathbb{P}(M_k|D) = \frac{BF[M_k:M_0]}{\sum_{l = 1}^K BF[M_l:M_0]}.
\end{equation}

\citet{Lenkoski2014} applied BMA to both the first and second stage of 2SLS in regression models without latent variables.  This procedure addresses uncertainty in both the selection of instruments as well as the combination of endogenous and exogenous predictors of the targeted outcome. This differs from applying BMA to a multiple regression model, as their approach averages over both first and second stage models, and weights the second stage coefficients by the product of both the first and second stage models' probability.  However, their combined approach is not appropriate in a SEM setting, as their approach performs the model averaging over both the first stage regression (corresponding to instruments) and the second stage regression (corresponding to the structural model). In a traditional instrumental variable regression setting, the instruments are auxillary to the predictor variables, making it possible to apply the model averaging in two stages as the set of instrumental variables is not determined by the selection of predictor variables.

In addition to Bayesian model averaging approaches applied to the use of instrumental variables, non-Bayesian model averaging approaches have been developed. Recently, \citet{sengStructuralEquationModel2022} apply least-squares model averaging to 2SLS estimation, and establish the asymptotic consistency of the least-squares weighting scheme. This work is extended in \citet{sengInstrumentalVariableModel2023}, detailing an approach for applying 2SLS estimation to high dimensional data by averaging first stage estimates of the endogenous variables, where weights are the estimated least squares regression coefficients of original endogenous variables regressed on all first stage estimates of said endogenous variables. While the authors apply their model averaging method to only the problem of instrument selection, the least squares model averaging approach could also be extended to variable selection at the second stage similar to what is done in \citet{Lenkoski2014}.

Regardless of the model averaging method used, the MIIVSEM setting presents a number of issues stemming from the \textit{model implied} nature of the instruments. In a MIIVSEM setting, the structural model and MIIV set are dependent on one another. As SEM models are multivariate in nature, iterating over all possible structural models for even a small number of variables becomes computationally intractable quickly. Additionally, each change in the endogenous predictor set of a given equation will change the possible MIIV set for that, and likely other equations. This then leads to non-comparable models for each equation as we iterate through all possible combinations of relations, making BMA at the instrument level intractable. Both because of these issues and the focus of this manuscript on expanding the capabilities of the MIIVSEM estimator for \textit{a priori} specified models, we restrict our approach only to the first stage of the estimator.

\subsection{Prior Choice}

We use the BMA approach to \textit{instrument selection}, as it can account for the uncertainty previously described regarding the specific MIIV set to use for any given equation. This approach still diverges from a simple application of BMA to multiple regression as here we are iterating over every instrument set and using the probability of the first stage model to weight the estimates from the second stage. Furthermore, our application of BMA improves on the approach of \citet{Lenkoski2014} by using univariate and multivariate empirical $g$-priors \citep{Zellner1986,Liang2008, Brown1998}: uninformative variable selection priors that lead to posterior mean estimates being asymptotically equivalent to the maximum likelihood point estimate of the parameter. The use of these empirical g-priors remove the need for prior hyperparameter elicitation, and allow researchers to use this method without the need for a sensitivity analysis.

Bayesian model averaging requires the choice of a prior on the parameters of interest. In a multiple linear regression framework with normally distributed errors, a common family of priors is that of the $g$-prior \citep{Zellner1986}. Given the standard linear regression model:

\begin{align*}
Y &= \mathbf{X}\boldsymbol{\beta} + \epsilon\\
\epsilon &\sim N(0, \sigma^2)
\end{align*}
the definition of a g-prior is then,
\begin{align*}
\mathcal{P}(\sigma) &\propto \frac{1}{\sigma^2}\\
\boldsymbol{\beta}|\sigma &\sim N\left(\tilde{\boldsymbol{\beta}}, g\sigma^2(\mathbf{X}^\prime\mathbf{X})^{-1}\right)
\end{align*}
where $g$ is chosen according to a variety of strategies that are discussed below. $\tilde{\boldsymbol{\beta}}$ is the prior mean for $\boldsymbol{\beta}$, and is usually set to $0$. The $\mathcal{P}(\sigma) \propto \frac{1}{\sigma^2}$ assumption corresponds to the Jeffrey's prior for a normal distribution with unknown variance \citep{Jeffreys1946}. The Jeffrey's prior is common choice for an uninformative  prior, as it depends only on the observed data.

The $g$-prior is particularly attractive for model selection and model averaging purposes, as the Bayes factor comparing any model to the null model is analytically defined \citep[Eq. 6]{Liang2008}
\begin{equation}
\label{BFeq}
BF[M_k:M_0] = (1+g)^{(n-p_k-1)/2}[1+g(1-R^2_k)]^{-(n-1)/2}
\end{equation}
where $n$ is the sample size, $p_k$ is the number of predictors in model $M_k$, and $R^2_k$ is the R squared value for model $M_k$. This expression leads to more support for $M_k$ relative to the null model as $R^2_k$ increases, with larger values of $g$ leading to a greater rate of increase in support as $R^2_k$ increases. $p_k$ acts as a penalty on increasing the number of predictors.  

\citet{Lenkoski2014} used a Unit Information Prior \citep{Kass1995} where $g = n$, the sample size, and additionally center their prior on the OLS estimate of a given regression coefficient for the first stage models, and the 2SLS estimate for a given regression coefficient in the second stage models. This has the effect of centering the resulting posterior mean for the second stage coefficients on the 2SLS estimate for a given instrument set, and we adopt the same choice of centering. However, when used in model selection, the Unit Information Prior results in what is known as the information paradox \citep{Liang2008}. This paradox, briefly, says that under the Unit Information Prior and a number of others, the Bayes Factor comparing a given model $M_k$ to the null model approaches a constant as $R^2_k \rightarrow 1$. This is to say, as the evidence for a given model becomes overwhelming, the Bayes Factor comparing that model to the null model approaches a constant, rather than tending to infinity, which would be expected. In this manuscript, we use the local empirical Bayes prior \citep[Eq. 9]{Liang2008}, where for a given model $M_k$

\begin{equation}
\label{gprior}
g_k = max\{F_k - 1, 0\}  
\end{equation}
where $F_k$ is the $F$ statistic for $M_k$ and is calculated as $\frac{R^2_k/p_k}{(1- R^2_k)/(n-1-p_k)}$, with $R^2_k$ being the $R^2$ of model $M_k$, $p_k$ the number of predictors in $M_k$ and $n$ the sample size. 

This empirical Bayes prior maximizes the marginal likelihood in this setting, does not experience the information paradox and is also consistent for model selection and for prediction \citep{Liang2008}. These properties in addition to its computational ease of implementation make it a good choice for our purposes. 

The use of the empirical Bayes prior previously described is appropriate when there is a single endogenous predictor, such as a indicator that loads onto a single latent factor. However, to apply BMA to cases where there are more than one endogenous predictor in an equation, a prior for a multivariate linear regression (i.e. a linear regression where the outcome is multivariate), a multivariate g-prior must be used. We use the approach of \citet{Brown1998}. The standard multivariate normal regression model is used:

\begin{align}
    \mathbf{Y} &= \mathbf{1}\boldsymbol{\alpha}^\prime + \mathbf{X}\mathbf{B} + \mathbf{\mathcal{E}} \\
    \mathbf{\mathcal{E}} &\sim N_q\left( \mathbf{0}, \boldsymbol{\Sigma}\right)
\end{align}
where $\mathbf{Y}$ is an $n \times q$ matrix of outcome variables, $X$ is an $n \times p$ matrix of observed covariates, $\mathbf{B}$ is a $p \times q$ matrix of regression coefficients, and $\mathbf{\mathcal{E}}$ is the multivariate normal distributed error component with population covariance matrix $\boldsymbol{\Sigma}$.

\citet{Brown1998} propose the following as a prior distribution for $\mathbf{B}$ given $\boldsymbol{\Sigma}$ and a given model $M_k$:

\begin{equation}
    \pi(\boldsymbol{B} | \boldsymbol{\Sigma}, M_k) \propto |H_{M_k}|^{-q/2}|\boldsymbol{\Sigma}|^{-p_{M_k}/2}\exp\left[{-\frac{1}{2}\text{tr}(\mathbf{B}\boldsymbol{\Sigma}^{-1}\mathbf{B}^{-1})}\right]
    \label{eq:MVPrior}
\end{equation}
where the prior mean of $\mathbf{B}$ is $0$, $p_{M_k}$ is the number of independent variables in model $M_k$ and $H_{m_k}$ is a $p_{M_k} \times p_{M_k}$ matrix of the following form:
Bayesian Model Averaging for Model Implied Instrumental Variable Two Stage Least Squares Estimators
\begin{equation}
\mathbf{H}_{M_k} = g \times (\mathbf{X_{M_k}}^\prime\mathbf{X_{M_k}})^{-1}
\end{equation}
where $\mathbf{X}_{M_k}$ is the $ n \times p_{M_k}$ matrix of covariates for model $M_k$.

\citet{Brown1998} then derive the posterior distribution for a model $M_k$ given $Y$ and $X$:

\begin{equation}
    \pi(M_k | \mathbf{Y}, \mathbf{X}) \propto (1+g)^{-qp_{M_k}/2}(|\mathbf{Q}_{M_k}|)^{-(n + \delta + q -1)/2}\pi(M_k)
\end{equation}
where $\delta$ is a hyperparameter for the prior distribution of $\boldsymbol{\Sigma}$, and needs to be small for weak prior information. $\delta = 3$ is a suggested choice. $\mathbf{Q}_{M_k}$ is a $q \times q$ matrix

\begin{equation}
    \mathbf{Q}_{M_k} = k\mathbf{I}_q + \frac{g}{g+1}(\mathbf{Y}^\prime\mathbf{Y} - \mathbf{Y}^\prime\mathbf{X}_{M_k}(\mathbf{X}^\prime_{M_k}\mathbf{X}_{M_k})^{-1}\mathbf{X}^\prime_{M_k}\mathbf{Y}) + \mathbf{Y}^\prime\mathbf{Y}/(g+1).
\end{equation}

To derive the empirical Bayes estimate for $g$, we have to determine the maximum marginal likelihood with respect to $g$. We do not derive this analytically here, and instead determine the optimal $g$ numerically.

Numerically finding the optimal $g$ is feasible when the number of equations with more than one endogenous predictor is fairly small (e.g.  10 equations with multiple endogenous predictors adds approximately a minute to compute time, though this time is highly dependent on the overall complexity of the model as well). In practice, this optimization runs on the order of half a second to five seconds, making it a reasonable approach to determining the empirical Bayes prior in this specific setting. 

We now can move to describe the MIIV-2SBMA estimator, as well as BMA variants of the Sargan's $\chi^2$ test of overidentification. We propose the Instrument Specific Sargan's $\chi^2$ test of overidentification, which allows researchers to examine sources of model misspecification in a more fine grained fashion.

\section{MIIV-2SBMA}

Consider the latent to observed variable transformation of the structural equation model illustrated earlier but defined fully in the Supplementary Materials.

\begin{equation}
Y_j = \mathbf{X}_j\boldsymbol{\theta}_j + \mathbf{u}_j
\end{equation} where $Y$ are the outcomes, $\mathbf{X}$ is the set of endogenous predictors (for example, a latent variable that has been replaced with its scaling indicator), $\boldsymbol{\theta}$ is a matrix of coefficients, and $\mathbf{u}$ is the composite error term.  
The $j$ subscript indicates the equation in question. For simplicity, we assume here that $\mathbf{X}_j$ consists of a single endogenous predictor, however this approach extends to more than one endogenous predictor as well.

Lenkoski et al. \citeyear{Lenkoski2014} consider a single regression equation for their 2SBMA estimation.  We consider a system of equations where the dependent variable and covariates in each latent to observed variable equation can differ.  As previously described this leads to a set of $p$ model implied instrumental variables for the $j$th equation, $\mathbf{Z}_j$. We then construct a set of all $\binom{p}{l}$ combinations of the columns of $\mathbf{Z}_j$, where $l$ ranges from $q +1$ to $p$, where $q$ is the number of endogenous predictors in equation $j$. This set of subsets represents all possible combinations of MIIVs for a given equation $j$.  Denote a specific subset of $\mathbf{Z}_j$ as $\mathbf{Z}_{j,k}$. The number of subsets of $\mathbf{Z}_j$ corresponds to $K$ in Eq. 1. Note that while we use all possible subsets of $\mathbf{Z}_j$, it is possible to use specific subsets of the model implied instruments. However, as these are model implied instruments, rather than auxiliary instruments, there is less of a theoretical impetus to select specific subsets of instruments rather than take a all possible combination approach.

For each set $\mathbf{Z}_{j,k}$ the first stage model is as follows \citep[Eq. 7]{Lenkoski2014}:

\begin{equation}
\hat{\mathbf{X}}_{j,k} =  \mathbf{Z}_{j,k}(\mathbf{Z}_{j,k}^\prime\mathbf{Z}_{j,k})^{-1}\mathbf{Z}_{j,k}^\prime \mathbf{X}_j.
\end{equation}
where $\hat{\mathbf{X}}_{j,k}$ are the first stage estimates of the endogenous predictors $\mathbf{X}_j$ in equation $j$ for instrument set $k$. We obtain the $R^2$ for the first stage regression in the standard fashion

\begin{equation}
R^2_{\mathbf{X}_j} | \mathbf{Z}_{j,k} = 1- \frac{(\mathbf{X}_j-\hat{\mathbf{X}}_{j,k})^\prime(\mathbf{X}_j-\hat{\mathbf{X}}_{j,k})}{(\mathbf{X}_j-\bar{\mathbf{X}_j})^\prime(\mathbf{X}_j-\bar{\mathbf{X}_j})}
\end{equation}
where $\bar{\mathbf{X}}_j$ is the mean of $\mathbf{X}_j$. The Bayes Factor for the $j$th equation using the $k$th MIIV subset is then calculated from Equations \ref{BFeq} and \ref{gprior} , and we denote this as $\mathbf{BF}_{j,k}$.

Finally, the probability of any first stage model relative to any other evaluated first stage model is

\begin{equation}
\pi_{j,k} = \frac{\mathbf{BF}_{j,k}}{\sum_{l = 1}^K \mathbf{BF}_{j,l}}.
\end{equation}

Once we have our probabilities of the first stage regressions, we can use BMA to calculate an estimate of our second stage parameters across all evaluated MIIV subsets. The second stage point posterior means of $\boldsymbol{\theta}_j$ for a specific MIIV set $\mathbf{V}_{j,k}$ are calculated the same as the MIIV-2SLS estimate (given $\tilde{\boldsymbol{\theta}}_j = \hat{\boldsymbol{\theta}}_{j,(2SLS)}$ as previously specified), as such
\begin{equation}
\hat{\boldsymbol{\theta}}_{j,k} = (\hat{\mathbf{X}}_{j,k}^\prime\hat{\mathbf{X}}_{j,k})^{-1}\hat{\mathbf{X}}_{j,k}^\prime Y_j.
\end{equation}

The model averaged posterior mean (which we use as our point estimate) of $\hat{\boldsymbol{\theta}}_j$ are then the average of these specific posterior means, weighted by the model probability of each first stage equation $\pi_{j,k}$

\begin{equation}
\hat{\boldsymbol{\theta}}_j =\sum_{l = 1}^K \pi_{j,l}\boldsymbol{\hat{\theta}}_{j,l}. 
\end{equation}

While the model averaged variances of $\boldsymbol{\theta}_j$ are

\begin{equation}
\hat{\sigma}^2_{\boldsymbol\theta_j} = \sum_{l = 1}^K \pi_{j,l}\hat{\sigma}^2_{\boldsymbol{\theta}_{j,l}} + \sum_{l = 1}^K \pi_{j,l}(\boldsymbol{\hat{\theta}}_{j,l} - \boldsymbol{\hat{\theta}}_{j})^2
\end{equation}
where $\hat{\sigma}^2_{\boldsymbol{\theta}_{j,l}}$ are calculated as usual (i.e. \citet[Eq. 23]{Bollen1996}). Note that these expressions differ from the BMA estimates presented in \citet{Lenkoski2014} in that they do not average across second stage models. As described previously, this is because we do not employ BMA at the second stage, instead using \textit{a priori} specified models.

Our MIIV-2SBMA estimator accounts for weak instruments by down-weighting their contribution during the first stage regression. Crucially, this allows researchers to use all model implied instruments without making \textit{a priori} decisions as to which instruments to include, as truly weak instruments (where the association between instrument and regressor is asymptotically 0) will be removed from the estimation, while other weak instruments will have their contributions weighted in relation to the strength of their association with the regressor. The MIIV-2SBMA approach also obtains model probabilities for the stage 1 models, which in turn allows us to develop instrument specific tests of model missspecification.

\subsection{BMA Sargan's $\chi^2$ Tests}

\citet{Lenkoski2014} propose a model averaged variant of the classic Sargan's test. We first describe this test, and then propose a new Instrument Specific Sargan's Test.

Let $p_{j,k}$ be the Sargan's $p$-value calculated in the standard fashion for the instrument subset $Z_{j,k}$. The BMA Sargan's (BMA-S) $p$-value from \citet{Lenkoski2014} is then 

\begin{equation}
\label{BMASp}
p_j = \sum_{k = 1}^K \pi_{j,k}p_{j,k}.
\end{equation}

While this test does take advantage of the model averaging over instrumental variables, it still lacks the ability to pinpoint which instrument is the source of model misspecification. Additionally, as is shown in the later simulation studies, averaging over all possible instrument sets \textit{reduces} the power of the Sargan's Test to detect single invalid instruments, as the majority of sets of MIIVs will not contain that instrument. However,  we can extend this test\footnote{The test as originally proposed by \citet{Lenkoski2014} also sums across the probabilities of the second stage models. As we fix the structure of our second stage models, this additional summation does not appear in Eq \ref{BMASp}.}
 to an \textit{Instrument Specific} Sargan's Test, which both detects specific invalid instruments as well as avoids the same decrease in power the BMA Sargan's Test has. Let the index set $Q$ denote the subsets of $Z_j$ that include a specific instrument, denoted $q$. 

We calculate the probability of the first stage model dependent on the presence of a specific instrument as follows:

\begin{equation}
\pi_{j,k}^{(q)} = \frac{BF_{j,k}}{\sum_{l \in Q}{BF_{j,l}}}
\end{equation}
where $k \in Q$. This renormalizes our posterior model probabilities to be over all models which use instrument $q$. Similarly, we collect the Sargan's $p$-values for our instrument sets that contain a specific variable as $p_{j,k}^{(q)}$. The Instrument Specific Sargan's $p$-value is then

\begin{equation}
\label{ISSargan}
p_{j}^{(q)} = \sum_{l \in Q} \pi_{j,l}^{(q)} p_{j,l}^{(q)}.
\end{equation}

This $p_{j}^{(q)}$ can be used to diagnose which MIIV is invalid, however it must be evaluated for every MIIV under consideration, and the relative differences between the $p$-values examined, as $p_{j}^{(q)}$ can be below the nominal $\alpha$ even if $q$ is a valid instrument. This can occur if there is a strong invalid instrument in the proposed MIIV set. Instead, the smallest $p_{j}^{(q)}$ is most likely to indicate the invalid instrument. For more detailed rationale behind the use of this minimum p-value heuristic, see Supplementary Materials.

\subsection{Weak Instrument Detection using Inclusion Probabilities}

Inclusion probabilities are calculated for each MIIV as follows. Let $Q$ again denote the index set for subsets of $Z_j$ that contain instrument $q$. The inclusion probability for $q$ is

\begin{equation}
\mathbb{P}(q)_j = \sum_{k \in Q} \pi_{j,k}.
\end{equation}

\citet{Lenkoski2014} note that these inclusion probabilities are direct measures of the weakness of specific instruments and suggest that instruments with lower inclusion probabilities $(\mathbb{P}(q)_j < .5)$ be dropped from the model. However, they also note that 2SBMA has the advantage of down-weighting the contribution of weak instruments, making the approach robust to their inclusion, which also is a property of MIIV-2SBMA. In the following simulation study, we evaluate the performance of MIIV-2SBMA in estimating parameters and detecting weak/invalid instruments under conditions of model misspecification.

It is also important to note that these inclusion probabilities reflect the weakness of instruments taking into account all other model implied instruments used. This means that an instrument might have a moderate or strong bivariate correlation with an endogenous variable, but is redundant with many other instruments, which would lead to a low inclusion probability. 

\section{Simulation Studies}
\tikzstyle{item}=[rectangle,draw=black,fill=white,
                    font=\scriptsize\sffamily\bfseries,inner sep=6pt]
\tikzstyle{factor}=[ellipse,draw=black,fill=white,
                     font=\bfseries,inner sep=6pt]
\tikzstyle{caption}=[font=\scriptsize\bfseries]
\tikzstyle{arr}=[-latex,black,line width=.5pt, font=\tiny, anchor=center]
\tikzstyle{doublearr}=[latex-latex,black,line width=.5pt, font=\scriptsize]
\tikzstyle{labBox}=[rectangle, draw=black, fill=white,inner sep = .8pt]

\begin{figure}[H]
\centering
\begin{tikzpicture}[scale=.8, show background rectangle, node distance=1cm, auto,]
\node[factor](f1) at (-3.5, 0) {$\eta_1$} ;
\node[factor](f2) at (1.5, 0) {$\eta_2$} ;
\node[item](y1) at (-5,-2) {$Y_1$};
\node[item](y2) at (-4,-2) {$Y_2$};
\node[item](y3) at (-3,-2) {$Y_3$};
\node[item](y4) at (-2,-2) {$Y_4$};
\node[item](y5) at (0,-2) {$Y_5$};
\node[item](y6) at (1,-2) {$Y_6$};
\node[item](y7) at (2,-2) {$Y_7$};
\node[item](y8) at (3,-2) {$Y_8$};
\draw[arr](f1) to node[labBox] {$1$} (y1.north) ;
\draw[arr](f1) to node[labBox] {$1$} (y2.north) ;
\draw[arr](f1) to node[labBox] {$1$} (y3.north);
\draw[arr](f1) to node[labBox] {$1$} (y4.north);
\draw[arr](f2) to node[labBox] {$1$} (y5.north);
\draw[arr](f2) to node[labBox] {$1$} (y6.north);
\draw[arr](f2) to node[labBox] {$1$} (y7.north);
\draw[arr](f2) to node[labBox] {$1$} (y8.north);
\draw[doublearr, bend left](f1.north) to node[anchor=center, above, font=\tiny] {FC} (f2.north);
\draw[doublearr, bend right, dashed](y2.south) to node[anchor=center, below, font=\tiny] {EC} (y3.south);
\node[caption](SIM1) at (-1, -3) {Simulation 1};

\node(var1)[font=\small, inner sep = .25pt] at (-4.5, 1) {1};
\node(var2)[font=\small, inner sep = .25pt] at (2.5, 1) {1};
\draw[arr](var1) to (f1);
\draw[arr](var2) to (f2);

\node(anchor1) at (-5,-3.5){};
\node(anchor2) at (3,-3.5){};

\draw[black, line width = .5](anchor1) to (anchor2);

\node[factor](f1b) at (-3.5, -6) {$\eta_1$} ;
\node[factor](f2b) at (1.5, -6) {$\eta_2$} ;
\node[item](y1b) at (-5,-8) {$Y_1$};
\node[item](y2b) at (-4,-8) {$Y_2$};
\node[item](y3b) at (-3,-8) {$Y_3$};
\node[item](y4b) at (-2,-8) {$Y_4$};
\node[item](y5b) at (0,-8) {$Y_5$};
\node[item](y6b) at (1,-8) {$Y_6$};
\node[item](y7b) at (2,-8) {$Y_7$};
\node[item](y8b) at (3,-8) {$Y_8$};
\draw[arr](f1b) to node[labBox] {$1$} (y1b.north) ;
\draw[arr](f1b) to node[labBox] {$1$} (y2b.north) ;
\draw[arr](f1b) to node[labBox] {$1$} (y3b.north);
\draw[arr](f1b) to node[labBox] {$1$} (y4b.north);
\draw[arr](f2b) to node[labBox] {$1$} (y5b.north);
\draw[arr](f2b) to node[labBox] {$1$} (y6b.north);
\draw[arr](f2b) to node[labBox] {$1$} (y7b.north);
\draw[arr](f2b) to node[labBox] {$1$} (y8b.north);
\draw[doublearr, bend left](f1b.north) to node[anchor=center, above, font=\tiny] {FC} (f2b.north);
\draw[doublearr, bend right, dashed](y2b.south) to node[anchor=center, below, font=\tiny] {EC} (y5b.south);
\node[caption](SIM2) at (-1, -9.75) {Simulation 2};
\node(var1b)[font=\small, inner sep = .25pt] at (-4.5, -5) {1};
\node(var2b)[font=\small, inner sep = .25pt] at (2.5, -5) {1};
\draw[arr](var1b) to (f1b);
\draw[arr](var2b) to (f2b);

\node(anchor3) at (-5,-10){};
\node(anchor4) at (3,-10){};

\draw[black, line width = .5](anchor3) to (anchor4);

\node[factor](f1c) at (-3.5, -16) {$\eta_1$} ;
\node[factor](f2c) at (1.5, -16) {$\eta_2$} ;
\node[factor](f3c) at (-1, -13) {$\eta_3$} ;
\node[item](y1c) at (-5,-18) {$Y_1$};
\node[item](y2c) at (-4,-18) {$Y_2$};
\node[item](y3c) at (-3,-18) {$Y_3$};
\node[item](y4c) at (-2,-18) {$Y_4$};
\node[item](y5c) at (0,-18) {$Y_5$};
\node[item](y6c) at (1,-18) {$Y_6$};
\node[item](y7c) at (2,-18) {$Y_7$};
\node[item](y8c) at (3,-18) {$Y_8$};
\node[item](y9c) at (-2.5,-11) {$Y_9$};
\node[item](y10c) at (-1.5,-11) {$Y_{10}$};
\node[item](y11c) at (-.5,-11) {$Y_{11}$};
\node[item](y12c) at (.5,-11) {$Y_{12}$};

\draw[arr](f1c) to node[labBox] {$1$} (y1c.north) ;
\draw[arr](f1c) to node[labBox] {$1$} (y2c.north) ;
\draw[arr](f1c) to node[labBox] {$1$} (y3c.north);
\draw[arr](f1c) to node[labBox] {$1$} (y4c.north);
\draw[arr](f2c) to node[labBox] {$1$} (y5c.north);
\draw[arr](f2c) to node[labBox] {$1$} (y6c.north);
\draw[arr](f2c) to node[labBox] {$1$} (y7c.north);
\draw[arr](f2c) to node[labBox] {$1$} (y8c.north);
\draw[arr](f3c) to node[labBox] {$1$} (y9c.south);
\draw[arr](f3c) to node[labBox] {$1$} (y10c.south);
\draw[arr](f3c) to node[labBox] {$1$} (y11c.south);
\draw[arr](f3c) to node[labBox] {$1$} (y12c.south);

\draw[arr](f1c.north) to node[anchor=center, left, font=\tiny] {RC} (f3c);
\draw[arr](f2c.north) to node[anchor=center, right, font=\tiny] {RC} (f3c);
\draw[doublearr, bend left](f1c.north) to node[anchor=center, above, font=\tiny] {FC} (f2c.north);
\draw[doublearr, bend left, dashed](y2c) to node[anchor=center,left, font=\tiny] {EC} (y9c);
\node[caption](SIM3) at (-1, -19.75) {Simulation 3};
\node(var1c)[font=\small, inner sep = .25pt] at (-4.25, -15) {1};
\node(var2c)[font=\small, inner sep = .25pt] at (2.5, -15) {1};
\node(var3c)[font=\small, inner sep = .25pt] at (.5, -13) {1};
\draw[arr](var1c) to (f1c);
\draw[arr](var2c) to (f2c);
\draw[arr](var3c) to (f3c);

\end{tikzpicture}
\label{sims}
\caption{Path diagrams for Simulations 1 and 2. Solid and dashed lines represent parameters in the data generating model. Solid lines represent parameters estimated in the fitted models. EC: Error Correlation, FC: Factor Correlation, RC: Regression Coefficient. EC, FC, and RC are all parameters systematically varied in the simulation study.}
\end{figure}

Figure \ref{sims} presents the three population generating models that we use, labeled Simulations 1, 2 and 3. In all cases the true model includes both the solid and dashed lines.  The misspecified model omits the parameters represented by the dashed lines. Simulated data were generated from a multivariate normal with mean vector of $0$ and a covariance matrix implied by the population generating models. All data were simulated using the \texttt{lavaan} R package \citep{lavaan}.  For both Simulation 1 and 2, we are interested in estimating the factor loading for $Y_2$ in the latent to observed variable transformed equation, which corresponds to  
\begin{equation}
\label{targetEQ}
Y_2 = \lambda_2 Y_1 + U_{Y_2}
\end{equation}
where $\lambda_2$ is the factor loading for $Y_2$ which has the true value of $1$ and $U_{Y_2}$ is a normally distributed error. For the true model in Simulation 1 the valid MIIVs for the $Y_2$ equation (Eq. \ref{targetEQ}) are $Y_4, Y_5, Y_6, Y_7$ and $Y_8$, and for Simulation 2 the true model implies that $Y_3, Y_4, Y_6, Y_7$ and $Y_8$ are valid MIIVs. Simulation 1 is designed to evaluate how well our methods perform under differing degrees of instrument invalidity (as controlled by the magnitude of the omitted error correlation), while Simulation 2 is designed to evaluate how instrument invalidity interacts with instrument strength (as controlled by the magnitude of the between factor correlation)

For Simulation 3, we are interested in estimating the regression coefficients predicting $\eta_3$ by $\eta_1$ and $\eta_2$. In the latent to observed variable framework this corresponds to 
\begin{equation}
    Y_9 = \beta_1Y_1 + \beta_2Y_5 + U_{Y_9}
\end{equation}
where $\beta_1 = \beta_2$ are regression coefficients, and are set to either .5 or 2. Simulation 3 was designed to evaluate the performance of the multivariate prior and $g$ parameter selection procedure (See Supplementary Materials for more information).

In Simulation $1$, the proposed model omits the error covariance between $Y_3$ and $Y_2$, while in Simulation $2$, the proposed model omits the error covariance between $Y_5$ and $Y_2$. Finally, in Simulation $3$, the proposed model omits the error covariance between $Y_9$ and $Y_2$. Omitting these error covariances lead to $Y_3$ mistakenly being included in the MIIV set for Simulation $1$, $Y_5$ mistakenly being included in the MIIV set for Simulation $2$, and $Y_2$ mistakenly being included in the MIIV set for Simulation $3$. They are invalid instruments for each simulation respectively. 

For both simulations, we set the inter-factor correlation (FC) at either $.1$  or $.8$ and vary the value of the omitted error covariance (EC) at values of $.1$ and $.6$. This full cross of conditions allows us to evaluate the impact of weak instruments and invalid instruments. Finally, all indicators had a total variance of $1$.

When the FC is .1, indicators $Y_5, Y_6, Y_7, Y_8$ are relatively weak instruments for Equation \ref{targetEQ}. $Y_3$ is an invalid instrument in Simulation $1$,  $Y_5$ is an invalid instrument in Simulation 2 and $Y_2$ is an invalid instrument in Simulation 3. We evaluate all simulations and conditions at a sample size of $100$ and $500$. For every combination of conditions, we ran $500$ replications.

\subsection{Estimators}

We examine the performance of MIIV-2SBMA and associated over-identification tests relative to the performance of two MIIV-2SLS estimates. The first comparison estimate we title ``Invalid MIIVs,'' and is the MIIV-2SLS estimate (and associated Sargan's test) that utilizes \textit{all} MIIVs, including the invalid ones. The second comparison estimate we title ``Correct MIIVs'', which is the MIIV-2SLS estimate that includes only MIIVs implied by the true model. In other words, for Simulation 1 the true MIIV set is $Y_4, Y_5, Y_6, Y_7, Y_8$, Simulation 2 $Y_3, Y_4, Y_6, Y_7, Y_8$ and for Simulation 3 $Y_3, Y_4, Y_6, Y_7, Y_8$. Our ``Correct MIIVs'' estimator allows us to examine the performance of the 2SLS estimator when invalid instruments are excluded. The ``Invalid MIIVs'' and ``Correct MIIVs'' estimates were calculated using the \texttt{MIIVsem} R package \citep{miivsem}. MIIV-2SBMA estimates were calculated using code available in the Supplementary Materials. Additionally, code to replicate these simulations is available in the Supplementary Materials.

\clearpage

\subsection{Outcomes}
Our outcomes of interest are as follows:
\begin{itemize}
\setlength\itemsep{1em}
\item Power of Sargan's Test: Rate of rejection of the Sargan's Test (traditional or BMA) at $\alpha = .05$ .
\item Power of the Instrument Specific Sargan's Test (MIIV-2SBMA only): Instrument wise rate of rejection of the Instrument Specific Sargan's Test (Eq. \ref{ISSargan}) at $\alpha = .05$.
\item Specificity of the Instrument Specific Sargan's Test: Proportion of replications in which a given instrument has the lowest Instrument Specific Sargan's Test $p$-value.
\item Average Inclusion Probability (MIIV-2SBMA only).
\end{itemize}
We also assess bias and absolute bias of the outcomes of interest (for Simulation 1 and 2, $\hat{\lambda}_2$, for Simulation 3, $\hat{\beta}_1$, $\hat{\beta}_2$) as well as the standard error of the estimate, $SE(\hat{\theta})$. Absolute and relative bias were comparable between the Invalid MIIVs estimator and the MIIV-2SBMA estimator, while bias of the Correct MIIVs estimator is, as expected, less than the other two estimators. We found that all three estimators had similar standard errors, with MIIV-2SBMA having very slightly increased standard errors at $N = 100$. The bias and standard error results are in the Supplementary Materials.

\section{Results}

\subsection{Simulation 1}

\begin{table}[H]
\centering
\caption{Simulation 1 Results: Sargan's Test Power. For Invalid and Correct MIIVs, Sargan's Test Power is for the traditional test. For MIIV-2SBMA, power is for the BMA Sargan's Test.}
\label{Sim1ResSargPower}

\begin{tabular}{lrcccccccc}
\hline
Model                                    &  \multicolumn{2}{c}{\begin{tabular}[c]{@{}c@{}}EC = .1\\ FC = .1\end{tabular}} & \multicolumn{2}{c}{\begin{tabular}[c]{@{}c@{}}EC = .6\\ FC = .1\end{tabular}} & \multicolumn{2}{c}{\begin{tabular}[c]{@{}c@{}}EC = .1\\ FC = .8\end{tabular}} & \multicolumn{2}{c}{\begin{tabular}[c]{@{}c@{}}EC = .6\\ FC = .8\end{tabular}} \\ \hline
Sample Size                              & 100                                   & 500                                   & 100                                   & 500                                   & 100                                   & 500                                   & 100                                   & 500                                   \\  \hline
                                     Invalid MIIVs            & 0.05                                  & 0.12                                  & 0.51                                  & 1                                     & 0.08                                  & 0.15                                  & 0.69                                  & 1                                     \\
 Correct MIIVs           & 0.05                                  & 0.04                                  & 0.07                                  & 0.05                                  & 0.06                                  & 0.06                                  & 0.05                                  & 0.05                                  \\
                                          MIIV-2SBMA                  & 0.01                                  & 0.15                                  & 0.41                                  & 1                                     & 0.03                                  & 0.15                                  & 0.27                                  & 1                                     \\ \hline
\end{tabular}

\end{table}

Recall that the goal of Simulation 1 is to estimate the loading for variable $y_2$ conditional on an omitted error covariance (EC) between variable $y_2$ and $y_3$. This misspecification renders $y_3$ an invalid instrument for the estimation of $y_2$'s loading.

Table \ref{Sim1ResSargPower} shows the power of traditional and BMA Sargan's tests across our three estimators. As expected, the Correct MIIVs estimator Sargan's Test power was close to the nominal $\alpha$ level of $.05$. For the Invalid MIIVs estimator, Sargan's test power exhibited the expected behavior, in that the power to detect invalid instruments increased with magnitude of the omitted error covariance (EC = .1 vs. EC = .6), and with increasing sample size. The power of BMA Sargan's test to detect invalid instruments was considerably below that of the Invalid MIIVs estimator (e.g. for EC = .6, FC = .6, MIIV-2SBMA power = .27 at n = 100, while Invalid MIIVs power = .69).

In light of these Sargan's Test power results, the Instrument Specific Sargan's Test needs to be evaluated. Figure \ref{VarSpecSargPSim1} and Table \ref{Sim1VarSpecPower} below show the power of the Instrument Specific Sargan's Test along with the instrument-wise specificity.

\begin{table}[H]
\centering
\caption{Simulation 1: Instrument Specific Sargan's Test Power with $\alpha$ set at $.05$. (Specificity). $y_3$ is the invalid instrument.}
\label{Sim1VarSpecPower}
\resizebox{\textwidth}{!}{%
\begin{tabular}{rcccccccc}
\hline
Model       & \multicolumn{2}{c}{\begin{tabular}[c]{@{}c@{}}EC = .1\\ FC = .1\end{tabular}} & \multicolumn{2}{c}{\begin{tabular}[c]{@{}c@{}}EC = .6\\ FC = .1\end{tabular}} & \multicolumn{2}{c}{\begin{tabular}[c]{@{}c@{}}EC = .1\\ FC = .8\end{tabular}} & \multicolumn{2}{c}{\begin{tabular}[c]{@{}c@{}}EC = .6\\ FC = .8\end{tabular}} \\ \hline
Sample Size & 100                                   & 500                                   & 100                                    & 500                                  & 100                                   & 500                                   & 100                                    & 500                                  \\ \hline
$Y_3$          & 0.04 (.17)                                 & 0.15 (.224)                                 & 0.57 (.442)                                  & 1 (.436)                                   & 0.07 (.232)                                  & 0.15 (.124)                                 & 0.79 (.746)                                  & 1 (.786)                                    \\
$Y_4$          & 0.02 (.088)                                 & 0.15 (.118)                                 & 0.45 (.236)                                  & 1 (.354)                                   & 0.03 (.09)                                 & 0.15  (.042)                                & 0.26   (.06)                                & 1   (.01)                                 \\
$Y_5$          & 0.02 (.198)                                 & 0.13 (.178)                                 & 0.35 (.104)                                  & 1 (.084)                                   & 0.04 (.188)                                 & 0.16 (.2)                                 & 0.28   (.044)                                & 1  (.056)                                  \\
$Y_6$          & 0.03 (.194)                                 & 0.12 (.17)                                 & 0.36  (.068)                                 & 1  (.032)                                  & 0.04  (.15)                                & 0.15  (.206)                                & 0.26   (.044)                                & 1  (.042)                                  \\
$Y_7$          & 0.03 (.216)                                & 0.14  (.148)                                & 0.36   (.086)                                & 1   (.056)                                 & 0.03  (.158)                                & 0.15 (.198)                                 & 0.25  (.056)                                 & 1  (.06)                                  \\
$Y_8$          & 0.03 (.134)                                 & 0.13 (.162)                                 & 0.37  (.064)                                 & 1  (.038)                                  & 0.05  (.182)                                & 0.16 (.23)                                 & 0.26  (.05)                                 & 1 (.046)                                   \\ \hline
\end{tabular}
}
\end{table}

\begin{figure}[H]
\centering
\includegraphics[width = .8\textwidth]{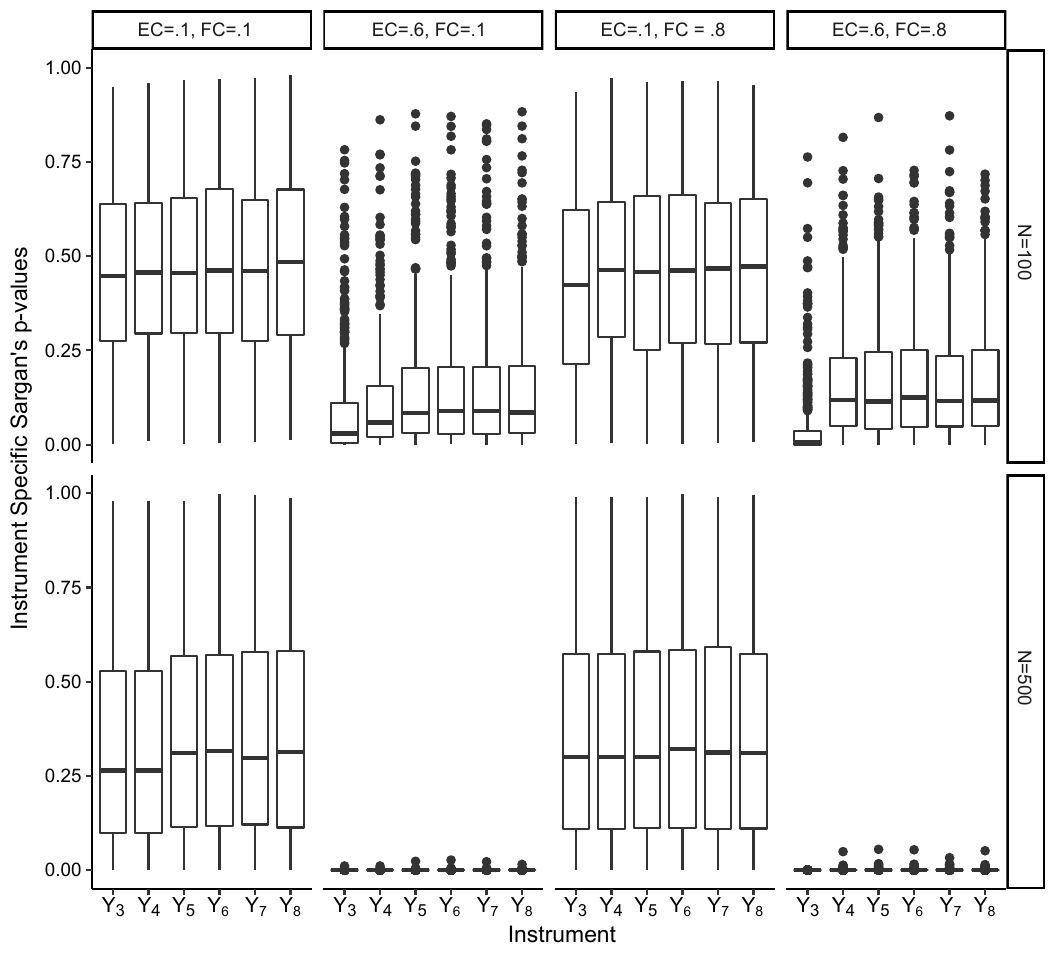}
\caption{Simulation 1: Instrument Specific Sargan's $p$-values. Box plots represent interquartile range. Black bar is the median. Whiskers indicate $1.5*IQR$. EC is the value of the error covariance, while FC is the value of the between factor covariance. N is the sample size.}
\label{VarSpecSargPSim1}
\end{figure}

The Instrument Specific Sargan's Test shows better power than the BMA Sargan's Test and the traditional Sargan's Test. Specifically, the power to detect $Y_3$ (which in Simulation 1 is the invalid IV) is greater in low sample sizes than in the traditional Sargan's Test from the Invalid MIIV estimator (.57 vs. .51 for the EC = .6, FC = .1 condition; .79 vs .69 for the EC = .6, FC = .8 condition). The pattern of the Instrument Specific Sargan's Tests also sheds light on the underpowered nature of the BMA Sargan's Test presented in Table \ref{Sim1ResSargPower}. That test averages Sargan's $p$-values over all instrument subsets, which includes instrument subsets that do not contain $Y_3$. Figure \ref{VarSpecSargPSim1} shows the property of the Instrument Specific Sargan's Test, in that while the $p$-value is, on average, low for all instruments, it is, on average, lowest for the invalid instrument $Y_3$. 

The proportion of replications in which a given instrument had the lowest $p$-value (Specificity; Table \ref{Sim1VarSpecPower}) further informs the ability of the Instrument Specific Sargan's Test to identify the invalid instrument. For conditions with high error covariances (EC = .6), $Y_3$ tended to have the minimum $p$-value. In the case of low between factor correlation (EC = .6, FC = .1), the proportion of times $Y_3$ had the minimum $p$-value was maximal, but relatively low at $.442$ for $N = 100$ and $.436$ for $N = 500$. Interestingly, in the same condition when $N = 500$, the proportion for $Y_4$ increases to $.354$ from $.236$ when $N = 100$. This condition (EC = .6, FC = .1) corresponds to the situation where there is a invalid instrument within the same factor as the target equation, while the indicators of the other factor are weak instruments. This result suggests that the Instrument Specific Sargan's Test is less specific in this particular case, and can localize the invalid instrument down to a specific factor structure rather than pinpointing the exact instrument. This is not the case when EC = .6 and FC = .8, where the indicators of the second factor ($Y_5$, $Y_6$, $Y_7$, $Y_8$) are stronger instruments when estimating $\lambda_2$. Here, the Instrument Specific Sargan's Test has better specificity in pinpointing $Y_3$ as the invalid instrument (.746, .786, $N = 100$ and $N = 500$ respectively). In the conditions where $Y_3$ has a small covariance with the equation error (EC = .1), the Instrument Specific Sargan's Test has both low power generally, and low specificity to detect $Y_3$ as the invalid instrument. 

Finally, we can examine our measure of weak instruments, the inclusion probabilities. Table \ref{Sim1IncProb} show the inclusion probabilities of each model implied instrument.

\begin{table}[H]
\centering
\captionsetup{justification=centering}
\caption{Simulation 1: Mean Inclusion Probabilities.}
\label{Sim1IncProb}
\begin{tabular}{rcccccccc}
\hline
Model & \multicolumn{2}{c}{\begin{tabular}[c]{@{}c@{}}EC = .1\\ FC = .1\end{tabular}} & \multicolumn{2}{c}{\begin{tabular}[c]{@{}c@{}}EC = .6\\ FC = .1\end{tabular}} & \multicolumn{2}{c}{\begin{tabular}[c]{@{}c@{}}EC = .1\\ FC = .8\end{tabular}} & \multicolumn{2}{c}{\begin{tabular}[c]{@{}c@{}}EC = .6\\ FC = .8\end{tabular}} \\ \hline
Sample Size  & 100                                   & 500                                   & 100                                   & 500                                   & 100                                   & 500                                   & 100                                   & 500                                   \\ \hline
$Y_3$    & 0.93                                  & 1                                     & 0.92                                  & 1                                     & 0.79                                  & 1                                     & 0.77                                  & 1                                     \\
$Y_4$    & 0.92                                  & 1                                     & 0.93                                  & 1                                     & 0.79                                  & 1                                     & 0.8                                   & 1                                     \\
$Y_5$    & 0.31                                  & 0.18                                  & 0.32                                  & 0.18                                  & 0.41                                  & 0.57                                  & 0.45                                  & 0.57                                  \\
$Y_6$    & 0.3                                   & 0.17                                  & 0.32                                  & 0.17                                  & 0.41                                  & 0.56                                  & 0.42                                  & 0.57                                  \\
$Y_7$    & 0.31                                  & 0.17                                  & 0.32                                  & 0.17                                  & 0.44                                  & 0.55                                  & 0.43                                  & 0.56                                  \\
$Y_8$    & 0.3                                   & 0.19                                  & 0.31                                  & 0.17                                  & 0.41                                  & 0.58                                  & 0.43                                  & 0.56                                  \\ \hline
\end{tabular}
\end{table}

The inclusion probabilities indicate for all conditions that $Y_3$ and $Y_4$ are strong instruments, which is to be expected as they are indicators of the same factor as $Y_1$ and $Y_2$. Here, it is important to note that inclusion probabilities do not account for possibly invalid instruments, as invalid instruments can be strongly related to the endogenous predictor. For instruments that are indicators of the second latent factor, the inclusion probabilities are dependent on the inter-factor correlation, with these instruments having higher inclusion probabilities with greater inter-factor correlations.

\subsection{Simulation 2 }

Recall that the purpose of Simulation 2 was to test how well the BMA Sargan's Test and Instrument Specific Sargan's Test can detect invalid and \textit{weak} instruments. As the invalid instrument for $\lambda_2$ is $y_5$, the weakness/strength of $y_5$ as an instrument is controlled by the inter-factor correlation (FC).
Power of the Sargan's Tests for model misspecification are contained in Table \ref{Sim2ResSargPower}.
The power of the various Sargan's Tests reveals several interesting patterns. The Invalid MIIVs estimator exhibits the expected pattern of power, with the Sargan's Test indicating the presence of invalid instruments in conditions with a high error covariance, and the power of this test increases with sample size. For the BMA Sargan's Test however, the power is very low in all conditions. This suggests that the BMA Sargan's Test would be incapable of detecting invalid instruments. An examination of the Instrument Specific Sargan's Test power reveals why.

\begin{table}[H]
\centering
\caption{Simulation 2 Results: Sargan's Test Power. For Invalid and Correct MIIVs, Sargan's Test Power is for the traditional test. For MIIV-2SBMA, power is for the BMA Sargan's Test.}
\label{Sim2ResSargPower}
\begin{tabular}{rlllllllll}
\hline
\multicolumn{1}{l}{Model}               & \multicolumn{2}{c}{\begin{tabular}[c]{@{}c@{}}EC = .1\\ FC = .1\end{tabular}} & \multicolumn{2}{c}{\begin{tabular}[c]{@{}c@{}}EC = .6\\ FC = .1\end{tabular}} & \multicolumn{2}{c}{\begin{tabular}[c]{@{}c@{}}EC = .1\\ FC = .8\end{tabular}} & \multicolumn{2}{c}{\begin{tabular}[c]{@{}c@{}}EC = .6\\ FC = .8\end{tabular}} \\ \hline
Sample Size &   100                                   & 500                                   & 100                                   & 500                                   & 100                                   & 500                                   & 100                                   & 500                                   \\\hline
                                Invalid MIIVs  & 0.07                                  & 0.14                                  & 0.85                                  & 1                                     & 0.09                                  & 0.16                                  & 0.81                                  & 1                                     \\
 Correct MIIVs & 0.05                                  & 0.04                                  & 0.06                                  & 0.05                                  & 0.06                                  & 0.04                                  & 0.06                                  & 0.05                                  \\
MIIV-2SBMA        & 0.01                                  & 0.05                                  & 0.03                                  & 0.07                                  & 0.01                                  & 0.03                                  & 0.05                                  & 0.3                                   \\ \hline
\end{tabular}

\end{table}

\begin{figure}[H]
\centering
\includegraphics[width = .8\textwidth]{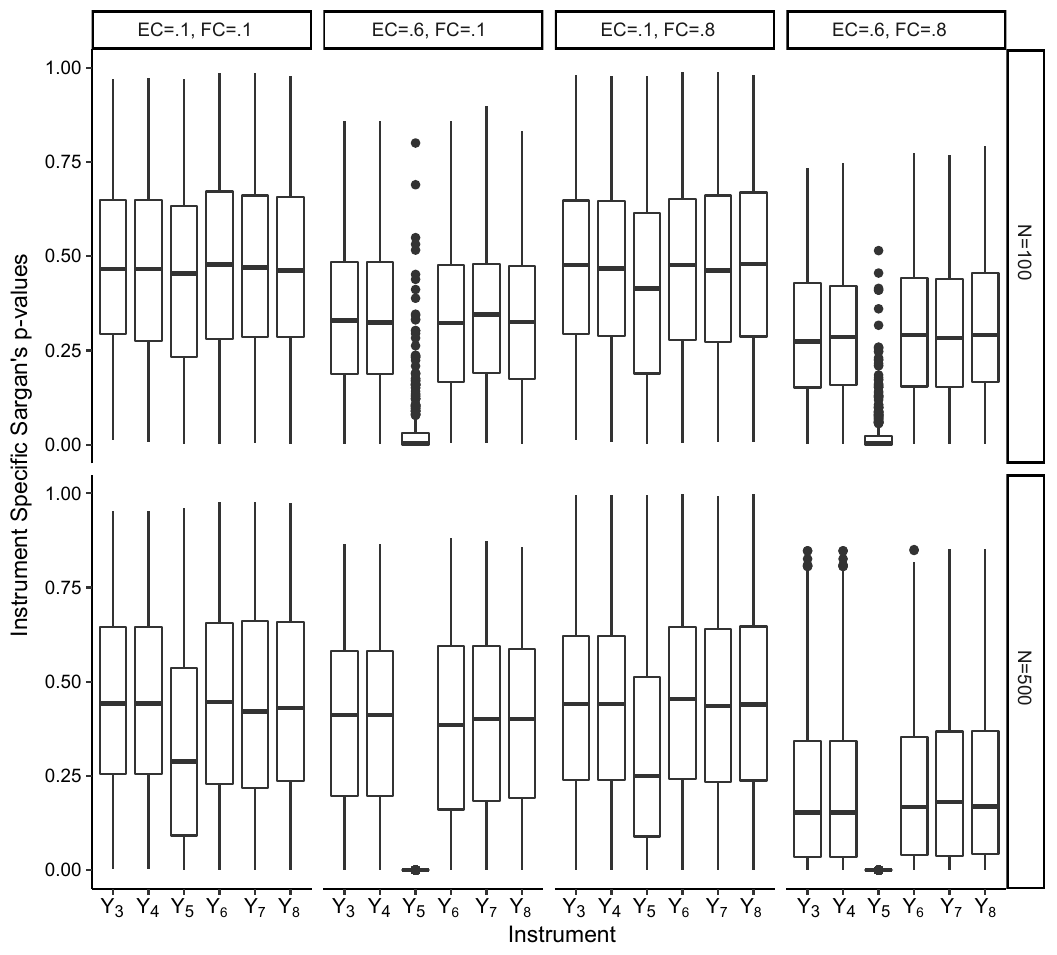}
\caption{Simulation 2: Instrument Specific Sargan's Test $p$-values. Box plots represent interquartile range. Black bar is the median. Whiskers indicate $1.5*IQR$. EC is the value of the error covariance, while FC is the value of the between factor covariance. N is the sample size.}
\label{VarSpecSargPSim2}
\end{figure}

Figure \ref{VarSpecSargPSim2} and Table \ref{VarSpecSargPSim2Table} show the Instrument Specific Sargan's Test $p$-value and power. $Y_5$ (which in Simulation 2 is invalid) is flagged with high probability as an invalid variable by the Instrument Specific Sargan's Test, while all other instruments are flagged at a much lower probability. This explains the low power of the BMA Sargan's Test, as it combines the $p$-values across all possible MIIV sets. The specificity results suggest that $Y_5$ can be identified as the invalid instrument with high probability for any condition. As one would expect, the specificity is relatively lower for $Y_5$ in conditions with low error covariance (EC = .1), and the specificity for $Y_5$ increases with sample size for each condition. Compared to Simulation 1, Simulation 2 suggests that the Instrument Specific Sargan's Test has good specificity for detecting weak invalid instruments.  Finally, we can examine inclusion probabilities.  

\begin{table}[H]
\centering
\caption{Simulation 2: Instrument Specific Sargan's Test Power with $\alpha$ set at $.05$. (Specificity). $y_5$ is the invalid instrument.}
\label{VarSpecSargPSim2Table}
\resizebox{\textwidth}{!}{%
\begin{tabular}{rcccccccc}
\hline
Model & \multicolumn{2}{c}{\begin{tabular}[c]{@{}c@{}}EC = .1\\ FC = .1\end{tabular}} & \multicolumn{2}{c}{\begin{tabular}[c]{@{}c@{}}EC = .6\\ FC = .1\end{tabular}} & \multicolumn{2}{c}{\begin{tabular}[c]{@{}c@{}}EC = .1\\ FC = .8\end{tabular}} & \multicolumn{2}{c}{\begin{tabular}[c]{@{}c@{}}EC = .6\\ FC = .8\end{tabular}} \\ \hline
Sample Size  & 100                                   & 500                                   & 100                                   & 500                                   & 100                                   & 500                                   & 100                                   & 500                                   \\ \hline
$Y_3$    & 0.02 (.092)                                 & 0.05 (.052)                                 & 0.06 (.006)                                  & 0.07 (0)                                 & 0.02  (.092)                                & 0.03  (.032)                                 & 0.07 (.002)                                  & 0.3 (0)                                   \\
$Y_4$    & 0.02 (.108)                                 & 0.05 (.048)                                 & 0.06 (.002)                                & 0.07   (0)                               & 0.03  (.086)                                & 0.03  (.016)                                & 0.06  (.002)                                & 0.3 (0)                                   \\
$Y_5$    & 0.05 (.276)                                 & 0.14 (.468)                                 & 0.8 (.926)                                  & 1 (1)                                    & 0.07 (.312)                                 & 0.18 (.594)                                 & 0.84 (.95)                                 & 1   (1)                                  \\
$Y_6$    & 0.03 (.168)                                 & 0.04 (.162)                                 & 0.05 (.026)                                  & 0.09 (0)                                 & 0.03 (.174)                                 & 0.04  (.106)                                 & 0.07 (.02)                                  & 0.28  (0)                                \\
$Y_7$    & 0.02  (.196)                                & 0.04  (.15)                                & 0.06  (.014)                                & 0.08  (0)                                 & 0.04  (.176)                                & 0.04  (.01)                                & 0.05 (.01)                                 & 0.28 (0)                                  \\
$Y_8$    & 0.03  (.16)                                & 0.05  (.12)                                & 0.06   (.026)                               & 0.09 (0)                                 & 0.03  (.16)                                & 0.04   (.118)                               & 0.07 (.016)                                 & 0.27  (0)                                \\ \hline
\end{tabular}
}
\end{table}

Table \ref{IncProbSim2Tab} show the inclusion probabilities for each instrument across all conditions. Note that the general pattern of inclusion probabilities is similar to that found in Simulation 1, where $Y_3$ and $Y_4$ have, on average, high inclusion probabilities, while the rest of the model implied instruments have inclusion probabilities dependent on the inter-factor correlation. This also informs the previous finding of low power for the model averaged Sargan's Test. As $Y_5$ has low average inclusion probabilities, this has the effect of down-weighting significant Sargan's $p$-values, which leads to lower power for the model averaged Sargan's Test. The Instrument Specific Sargan's Test uses conditional probabilities in its weighting, which ignores the weakness of $Y_5$ as an instrument. This leads to the Instrument Specific Sargan's Test being optimal in detecting misspecification issues due to a specific instrument, particularly when that instrument is weak. 
\begin{table}[H]
\centering
\caption{Simulation 2: Mean Inclusion Probabilities}
\label{IncProbSim2Tab}
\begin{tabular}{rcccccccc}
\hline
Model & \multicolumn{2}{c}{\begin{tabular}[c]{@{}c@{}}EC = .1\\ FC = .1\end{tabular}} & \multicolumn{2}{c}{\begin{tabular}[c]{@{}c@{}}EC = .6\\ FC = .1\end{tabular}} & \multicolumn{2}{c}{\begin{tabular}[c]{@{}c@{}}EC = .1\\ FC = .8\end{tabular}} & \multicolumn{2}{c}{\begin{tabular}[c]{@{}c@{}}EC = .6\\ FC = .8\end{tabular}} \\ \hline
Sample Size  & 100                                   & 500                                   & 100                                   & 500                                   & 100                                   & 500                                   & 100                                   & 500                                   \\ \hline
$Y_3$    & 0.92                                  & 1                                     & 0.93                                  & 1                                     & 0.8                                   & 1                                     & 0.77                                  & 1                                     \\
$Y_4$    & 0.92                                  & 1                                     & 0.92                                  & 1                                     & 0.82                                  & 1                                     & 0.8                                   & 1                                     \\
$Y_5$    & 0.32                                  & 0.18                                  & 0.32                                  & 0.18                                  & 0.41                                  & 0.57                                  & 0.44                                  & 0.57                                  \\
$Y_6$    & 0.31                                  & 0.18                                  & 0.31                                  & 0.18                                  & 0.43                                  & 0.58                                  & 0.45                                  & 0.57                                  \\
$Y_7$    & 0.32                                  & 0.18                                  & 0.31                                  & 0.18                                  & 0.42                                  & 0.55                                  & 0.42                                  & 0.55                                  \\
$Y_8$    & 0.32                                  & 0.18                                  & 0.31                                  & 0.17                                  & 0.4                                   & 0.57                                  & 0.41                                  & 0.58                                  \\ \hline
\end{tabular}
\end{table}

\subsection{Simulation 3}

Simulation 3 was designed to test the ability of the multivariate prior specified in Eq \ref{eq:MVPrior} to correctly detect invalid instruments when the second stage equation consists of more than one endogenous predictor. Note that in this Simulation, $y_2$ is an invalid instrument for the desired regression, due to its error covariance with $y_9$. 

Table \ref{tab:Sim3SargPower} shows the power of the traditional Sargan's Test and the BMA Sargan's test to detect model misspecification.
The BMA Sargan's test exhibits reduced power to detect model misspecification in low sample size conditions (N = 100), while its power to detect misspecification is comparable to that of the traditional Sargan's test at larger sample sizes.

\begin{table}[H]
\centering
\caption{Simulation 3 Results: Sargan's Test Power. For Invalid and Correct MIIVs, Sargan's Test Power is for the traditional test. For MIIV-2SBMA, power is for the BMA Sargan's Test.}
\label{tab:Sim3SargPower}
\resizebox{\textwidth}{!}{%
\begin{tabular}{rrcccccccccccccccc}
\hline
Model       & \multicolumn{2}{c}{\begin{tabular}[c]{@{}c@{}}EC = .1\\ FC = .1\\ RC = .5\end{tabular}} & \multicolumn{2}{c}{\begin{tabular}[c]{@{}c@{}}EC = .1\\ FC = .1\\ RC = 2\end{tabular}} & \multicolumn{2}{c}{\begin{tabular}[c]{@{}c@{}}EC = .6\\ FC = .1\\ RC = .5\end{tabular}} & \multicolumn{2}{c}{\begin{tabular}[c]{@{}c@{}}EC = .6\\ FC = .1\\ RC = 2\end{tabular}} & \multicolumn{2}{c}{\begin{tabular}[c]{@{}c@{}}EC = .1\\ FC = .8\\ RC = .5\end{tabular}} & \multicolumn{2}{c}{\begin{tabular}[c]{@{}c@{}}EC = .1\\ FC = .8\\ RC = 2\end{tabular}} & \multicolumn{2}{c}{\begin{tabular}[c]{@{}c@{}}EC = .6\\ FC = .8\\ RC = .5\end{tabular}} & \multicolumn{2}{c}{\begin{tabular}[c]{@{}c@{}}EC = .6\\ FC = .8\\ RC = 2\end{tabular}} \\ \hline
Sample Size & 100                                         & 500                                       & 100                                        & 500                                       & 100                                         & 500                                       & 100                                        & 500                                       & 100                                         & 500                                       & 100                                        & 500                                       & 100                                        & 500                                        & 100                                        & 500                                       \\ \hline
Invalid MIIVs    & 0.06                                        & 0.12                                      & 0.05                                       & 0.05                                      & 0.64                                        & 1                                         & 0.19                                       & 0.77                                      & 0.04                                        & 0.11                                      & 0.06                                       & 0.08                                      & 0.55                                       & 1                                          & 0.17                                       & 0.75                                      \\
Correct MIIVs    & 0.05                                        & 0.07                                      & 0.05                                       & 0.04                                      & 0.03                                        & 0.04                                      & 0.05                                       & 0.06                                      & 0.05                                        & 0.04                                      & 0.05                                       & 0.05                                      & 0.03                                       & 0.04                                       & 0.05                                       & 0.05                                      \\
 MIIV-2SBMA       & 0.01                                        & 0.12                                      & 0                                          & 0.04                                      & 0.05                                        & 0.98                                      & 0.01                                       & 0.74                                      & 0                                           & 0.08                                      & 0                                          & 0.04                                      & 0.01                                       & 0.87                                       & 0.01                                       & 0.64                                      \\ \hline
\end{tabular}%
}
\end{table}

\begin{figure}[H]
\centering
\includegraphics[width = .99\textwidth]{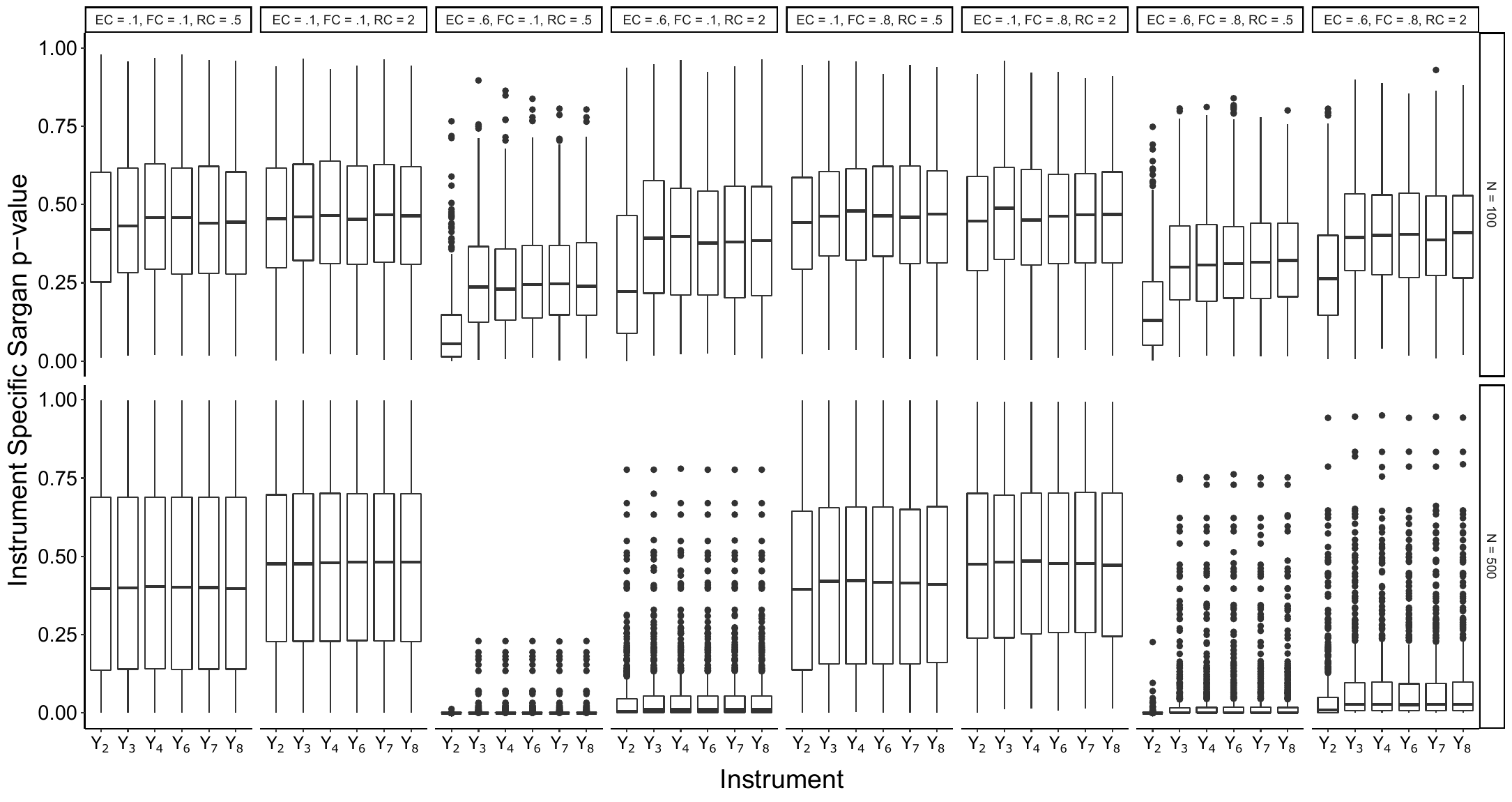}
\caption{Simulation 3: Instrument Specific Sargan's Test $p$-values. Box plots represent interquartile range. Black bar is the median. Whiskers indicate $1.5*IQR$. EC is the value of the error covariance, FC is the value of the between factor covariance, and RC is the true value of both inter-factor regression coefficients. N is the sample size. Note that $y_2$ is the invalid instrument in all conditions.}
\label{VarSpecSargPSim3}
\end{figure}

\begin{table}[H]
\centering
\caption{Simulation 3: Instrument Specific Sargan's Test Power with $\alpha$ set at $.05$. (Specificity). $y_2$ is the invalid instrument.}
\label{tab:Sim3VarSpecPart2}
\resizebox{\textwidth}{!}{%
\begin{tabular}{rcccccccc}
\hline
Model       & \multicolumn{2}{c}{\begin{tabular}[c]{@{}c@{}}EC = .1\\ FC = .8\\ RC = .5\end{tabular}} & \multicolumn{2}{c}{\begin{tabular}[c]{@{}c@{}}EC = .1\\ FC = .8\\ RC = 2\end{tabular}} & \multicolumn{2}{c}{\begin{tabular}[c]{@{}c@{}}EC = .6\\ FC = .8\\ RC = .5\end{tabular}} & \multicolumn{2}{c}{\begin{tabular}[c]{@{}c@{}}EC = .6\\ FC = .8\\ RC = 2\end{tabular}} \\ \hline
Sample Size & 100                                        & 500                                        & 100                                        & 500                                       & 100                                        & 500                                        & 100                                        & 500                                       \\ \hline
$y_2$       & 0.01 (0.216)                               & 0.12 (0.352)                               & 0.02 (0.218)                               & 0.06 (0.26)                               & 0.25 (0.659)                               & 0.99 (0.81)                                & 0.09 (0.502)                               & 0.75 (0.678)                              \\
$y_3$       & 0 (0.144)                                  & 0.08 (0.148)                               & 0.01 (0.124)                               & 0.04 (0.136)                              & 0.01 (0.072)                               & 0.87 (0.08)                                & 0.01 (0.076)                               & 0.64 (0.066)                              \\
$y_4$       & 0 (0.154)                                  & 0.09 (0.116)                               & 0.01 (0.182)                               & 0.04 (0.128)                              & 0.03 (0.096)                               & 0.87 (0.094)                               & 0.01 (0.092)                               & 0.63 (0.104)                              \\
$y_6$       & 0.01 (0.16)                                & 0.08 (0.112)                               & 0.01 (0.162)                               & 0.05 (0.158)                              & 0.01 (0.052)                               & 0.87 (0.006)                               & 0.02 (0.108)                               & 0.64 (0.05)                               \\
$y_7$       & 0 (0.188)                                  & 0.08 (0.154)                               & 0.01 (0.16)                                & 0.04 (0.148)                              & 0.02 (0.056)                               & 0.87 (0.002)                               & 0.01 (0.112)                               & 0.65 (0.062)                              \\
$y_8$       & 0.01 (0.138)                               & 0.08 (0.118)                               & 0.01 (0.154)                               & 0.04 (0.17)                               & 0.03 (0.064)                               & 0.87 (0.008)                               & 0.01 (0.11)                                & 0.63 (0.04)                               \\ \hline
\end{tabular}%
}
\end{table}

Fig. \ref{VarSpecSargPSim3} and Table \ref{tab:Sim3VarSpecPart2} show the results for the Instrument Specific Sargan's Test. These results show that the Instrument Specific Sargan's test exhibits high specificity in correctly determining which instrument is invalid based on having the smallest $p$-value. Overall power to detect model misspecification is lower at low sample sizes, but increases to more reasonable levels at N = 500. Specificity increases for the true invalid instrument ($y_2$) at higher sample sizes, and decreases for all other instruments at higher sample sizes.

\begin{table}[H]
\centering
\caption{Simulation 3: Mean Inclusion Probabilities}
\label{tab:Sim3IncProb}
\resizebox{\textwidth}{!}{%
\begin{tabular}{rcccccccccccccccc}
\hline
Model       & \multicolumn{2}{c}{\begin{tabular}[c]{@{}c@{}}EC = .1\\ FC = .1\\ RC = .5\end{tabular}} & \multicolumn{2}{c}{\begin{tabular}[c]{@{}c@{}}EC = .1\\ FC = .1\\ RC = 2\end{tabular}} & \multicolumn{2}{c}{\begin{tabular}[c]{@{}c@{}}EC = .6\\ FC = .1\\ RC = .5\end{tabular}} & \multicolumn{2}{c}{\begin{tabular}[c]{@{}c@{}}EC = .6\\ FC = .1\\ RC = 2\end{tabular}} & \multicolumn{2}{c}{\begin{tabular}[c]{@{}c@{}}EC = .1\\ FC = .8\\ RC = .5\end{tabular}} & \multicolumn{2}{c}{\begin{tabular}[c]{@{}c@{}}EC = .1\\ FC = .8\\ RC = 2\end{tabular}} & \multicolumn{2}{c}{\begin{tabular}[c]{@{}c@{}}EC = .6\\ FC = .8\\ RC = .5\end{tabular}} & \multicolumn{2}{c}{\begin{tabular}[c]{@{}c@{}}EC = .6\\ FC = .8\\ RC = 2\end{tabular}} \\ \hline
Sample Size & 100                                        & 500                                        & 100                                        & 500                                       & 100                                        & 500                                        & 100                                        & 500                                       & 100                                        & 500                                        & 100                                        & 500                                       & 100                                        & 500                                        & 100                                        & 500                                       \\ \hline
$Y_2$       & 0.6                                        & 0.99                                       & 0.61                                       & 0.99                                      & 0.6                                        & 0.99                                       & 0.63                                       & 0.99                                      & 0.53                                       & 0.95                                       & 0.53                                       & 0.95                                      & 0.54                                       & 0.94                                       & 0.54                                       & 0.96                                      \\
$Y_3$       & 0.61                                       & 0.99                                       & 0.61                                       & 0.99                                      & 0.6                                        & 0.99                                       & 0.61                                       & 0.99                                      & 0.54                                       & 0.95                                       & 0.54                                       & 0.95                                      & 0.54                                       & 0.96                                       & 0.53                                       & 0.96                                      \\
$Y_4$       & 0.61                                       & 0.99                                       & 0.59                                       & 0.99                                      & 0.63                                       & 0.99                                       & 0.6                                        & 0.99                                      & 0.52                                       & 0.96                                       & 0.56                                       & 0.96                                      & 0.54                                       & 0.95                                       & 0.52                                       & 0.94                                      \\
$Y_6$       & 0.6                                        & 0.99                                       & 0.62                                       & 0.99                                      & 0.62                                       & 0.99                                       & 0.62                                       & 0.99                                      & 0.55                                       & 0.96                                       & 0.53                                       & 0.94                                      & 0.53                                       & 0.96                                       & 0.52                                       & 0.96                                      \\
$y_7$       & 0.61                                       & 0.99                                       & 0.61                                       & 0.99                                      & 0.61                                       & 0.99                                       & 0.59                                       & 0.99                                      & 0.53                                       & 0.96                                       & 0.53                                       & 0.96                                      & 0.55                                       & 0.95                                       & 0.55                                       & 0.95                                      \\
$Y_8$       & 0.63                                       & 0.99                                       & 0.6                                        & 0.99                                      & 0.6                                        & 0.99                                       & 0.61                                       & 0.99                                      & 0.54                                       & 0.96                                       & 0.53                                       & 0.95                                      & 0.52                                       & 0.96                                       & 0.54                                       & 0.95                                      \\ \hline
\end{tabular}%
}
\end{table}

Table \ref{tab:Sim3IncProb} show the inclusion probabilities for all instruments. As no instrument was a weak one in this simulation condition, all inclusion probabilities are high across all conditions.

\section{Political Democracy Example}

\tikzstyle{item}=[rectangle,draw=black,fill=white,
                    font=\scriptsize\sffamily\bfseries,inner sep=6pt]
\tikzstyle{factor}=[ellipse,draw=black,fill=white,
                     font=\bfseries,inner sep=6pt]
\tikzstyle{caption}=[font=\scriptsize\bfseries]
\tikzstyle{arr}=[-latex,black,line width=.5pt, font=\tiny, anchor=center]
\tikzstyle{doublearr}=[latex-latex,black,line width=.5pt, font=\scriptsize]
\tikzstyle{labBox}=[rectangle, draw=black, fill=white,inner sep = .8pt]

\begin{figure}[h]
\centering
\begin{tikzpicture}[scale=.9, show background rectangle, node distance=1cm, auto,]
\node[factor](f1) at (-3.5, 0) {$\eta_1$} ;
\node[factor](f2) at (1.5, 0) {$\eta_2$} ;
\node[item](y1) at (-5,-2) {$Y_1$};
\node[item](y2) at (-4,-2) {$Y_2$};
\node[item](y3) at (-3,-2) {$Y_3$};
\node[item](y4) at (-2,-2) {$Y_4$};
\node[item](y5) at (0,-2) {$Y_5$};
\node[item](y6) at (1,-2) {$Y_6$};
\node[item](y7) at (2,-2) {$Y_7$};
\node[item](y8) at (3,-2) {$Y_8$};
\draw[arr](f1) to node[labBox] {$1$} (y1.north) ;
\draw[arr](f1) to node[labBox] {$\lambda_2$} (y2.north) ;
\draw[arr](f1) to node[labBox] {$\lambda_3$} (y3.north);
\draw[arr](f1) to node[labBox] {$\lambda_4$} (y4.north);
\draw[arr](f2) to node[labBox] {$1$} (y5.north);
\draw[arr](f2) to node[labBox] {$\lambda_6$} (y6.north);
\draw[arr](f2) to node[labBox] {$\lambda_7$} (y7.north);
\draw[arr](f2) to node[labBox] {$\lambda_8$} (y8.north);
\draw[doublearr, bend left](f1.north) to node[anchor=center, above, font=\tiny] {} (f2.north);
\draw[doublearr, bend right, dashed](y2.south) to node[anchor=center, below, font=\tiny] {} (y6.south);
\draw[doublearr, bend right, dashed](y6.south) to node[anchor=center, below, font=\tiny] {} (y8.south);
\draw[doublearr, bend right, dashed](y2.south) to node[anchor=center, below, font=\tiny] {} (y4.south);

\node[caption](SIM1) at (-1, -3.5) {Political Democracy};

\end{tikzpicture}
\label{PolDem2}
\caption{Path diagram for the Political Democracy model. Dashed lines represent error covariances in the original model that are relevant for the estimation of $\lambda_2$ and $\lambda_6$. Variance notation is suppressed. Additionally, error covariances that are not relevant to the estimation of $\lambda_2$ and $\lambda_6$. For more information about the complete structure, see \citet{Bollen1996}.}
\end{figure}

Our example examines the measurement structure of political democracy measures for 75 developing countries measured at 1960 and 1965. These data have been used previously to illustrate the MIIV-2SLS estimator \citep{Bollen1996} and thus makes for a useful dataset to compare the MIIV-2SLS estimator to the MIIV-2BMA estimator. These data are publicly available and included in the \texttt{MIIVsem} \citep{miivsem} and \texttt{laavan} \citep{lavaan} packages. Code to replicate this empirical example is available in the Supplementary Materials.

Figure 6 shows a CFA model with the latent political democracy variables correlated over time. $\eta_1$ is the latent factor for political democracy in 1960, while $\eta_2$ is the latent factor for political democracy in 1965. The original model has more correlated errors, but we have reduced the number for the sake of illustration. In Figure 6, the dashed lines represent error covariances from \citet{Bollen1996} that are relevant for the estimation of $\lambda_2$ and $\lambda_6$. We illustrate the use of MIIV-2SBMA by focusing on the factor loadings for $Y_2$ and $Y_6$, $\lambda_2$ and $\lambda_6$ respectively and examine how we can use the MIIV-2SBMA estimator to identify misspecifications. In our example below we omit the dashed error covariances to illustrate the effect of including invalid instruments. The choice to focus on these specific factor loadings was made to highlight two different cases, one where the presence of invalid instruments has a large impact on the estimate, and the other where the invalid instruments are also relatively weak and have less of an impact.  

In the original model, the MIIV set for $Y_2$ consists of $Y_3, Y_5, Y_7$ and $Y_8$. By omitting the error covariances between $Y_2$ and $Y_4$, and between $Y_2$ and $Y_6$, both $Y_4$ and $Y_6$ are added to the MIIV set. In this case, we can consider $Y_2$ and $Y_4$ as \textit{a priori} invalid instruments for the estimation of $\lambda_2$.

Similarly for $Y_6$, the original model leads to a MIIV set of $Y_1, Y_3, Y_4$ and $Y_7$, while omitting the error covariances between $Y_6$ and $Y_2$, and $Y_6$ and $Y_8$, leads to both $Y_2$ and $Y_8$ being included in the MIIV set. Here, we can consider $Y_2$ and $Y_8$ as \textit{a priori} invalid instruments for the estimation of $\lambda_6$.

The purpose of the empirical example is to illustrate how MIIV-2SBMA can identify the \textit{a priori} invalid instruments, and compare the factor loading estimates from the MIIV-2SLS and MIIV-2BMA approaches.

\subsection{Estimating $\lambda_2$ }

Table \ref{l2Table} shows estimates from MIIV-2SLS and MIIV-2SBMA, as well as Instrument Specific Sargan's Tests and inclusion probabilities for 3 stages of model building. Stage 1 uses all implied MIIVs, Stage 2 omits $Y_4$ as a MIIV, and Stage 3 omits both $Y_4$ and $Y_6$, therefore corresponding to the original model.

\begin{table}[H]
\centering
\caption{Estimates for $\lambda_2$. The lowest Instrument Specific Sargan's $p$ is bolded for each stage. Traditional Sargan's $p$-value (S $p$) and BMA Sargan's $p$-value (BMA-S $p$).}
\label{l2Table}
\resizebox{\textwidth}{!}{%
\begin{tabular}{c|ccc|ccc|cccccc|cccccc}
\hline
Stage   & \multicolumn{3}{c|}{MIIV-2SLS} & \multicolumn{3}{c|}{MIIV-2SBMA} & \multicolumn{6}{c|}{ Instrument Specific Sargan's $p$}                                     & \multicolumn{6}{c}{Inclusion Probabilities}                       \\
        & Estimate   & SE      & S $p$   & Estimate   & SE     & BMA-S $p$  & $Y_3$ & $Y_4$          & $Y_5$ & $Y_6$          & $Y_7$ & $Y_8$ & $Y_3$ & $Y_4$ & $Y_5$ & $Y_6$ & $Y_7$ & $Y_8$ \\\hline 
Stage 1 & 1.246      & 0.171   & 0.011   & 1.217      & 0.174  & 0.025     & 0.025 & \textbf{0.005} & 0.025 & 0.012          & 0.036 & 0.055 & 0.98  & 0.26  & 0.99  & 0.88  & 0.15  & 0.21  \\
Stage 2 & 1.216      & 0.171   & 0.047   & 1.208      & 0.173  & 0.032     & 0.032 & -              & 0.032 & \textbf{0.015} & 0.046 & 0.07  & 0.99  & -     & 0.99  & 0.99  & 0.15  & 0.21  \\
Stage 3 & 1.143      & 0.172   & 0.205   & 1.125      & 0.174  & 0.227     & 0.227 & -              & 0.227 & -              & 0.206 & 0.166 & 0.98  & -     & 0.99  & -     & 0.19  & 0.77  \\ \hline
\end{tabular}%
}
\end{table}

Focusing first on the results from the MIIV-2SLS estimator, we note that for both Stages 1 and 2, the traditional Sargan's Test (S $p$ in Table \ref{l2Table}) detects the presence of invalid instruments. Here, the Bayesian Model Averaged Sargan's Test (BMA-S $p$ in Table \ref{l2Table}) also detects the presence of invalid instruments at Stage 1 and 2. What is notable here is that the Instrument Specific Sargan's Test picks out $Y_4$ and $Y_6$ in Stage 1 and Stage 2 respectively as the invalid instruments, when one looks at the lowest Instrument Specific Sargan's $p$-value at each stage. The removal of $Y_4$ leads to a larger reduction in the estimate of $\lambda_2$ for the MIIV-2SLS estimator ($1.246$ to $1.216$, difference of $.03$) than for the MIIV-2SBMA estimator ($1.217$ to $1.208$, difference of $.009$). This can be explained by the low inclusion probability of $Y_4$ ($.26$), which suggests that $Y_4$'s contribution as an invalid instrument was down-weighted in the MIIV-2SBMA estimator. This can be confirmed by examining the change in the estimate when removing $Y_6$ from the MIIV set. In the MIIV-2SLS case, this leads to a $.073$ difference in the estimate, while in the MIIV-2SBMA case, this leads to a $.083$ difference in the estimate. The similar relative sizes of this difference in the estimate is due to $Y_6$ not only being an invalid instrument, but also a strong instrument, as its inclusion probability of $.99$ suggests. As it is a strong instrument, its biasing impact is not down-weighted by MIIV-2SBMA. As per the results from the simulation study, this suggests that MIIV-2SBMA improves upon MIIV-2SLS by being more robust to weak, invalid instruments.

\subsection{Estimating $\lambda_6$}

Table \ref{l6Table} shows estimates from MIIV-2SLS and MIIV-2SBMA, as well as Instrument Specific Sargan's Tests and inclusion probabilities for 3 stages of model building. Stage 1 uses all implied MIIVs, Stage 2 omits $Y_4$ as a MIIV, and Stage 3 omits both $Y_4$ and $Y_8$, therefore corresponding to the original model.

\begin{table}[H]
\centering
\caption{Estimates for $\lambda_6$. The lowest Instrument Specific Sargan's $p$ is bolded for each stage. Traditional Sargan's $p$-value (S $p$) and BMA Sargan's $p$-value (BMA-S $p$). }
\label{l6Table}
\resizebox{\textwidth}{!}{%
\begin{tabular}{c|ccc|ccc|cccccc|cccccc}
\hline
Stage   & \multicolumn{3}{c|}{MIIV-2SLS} & \multicolumn{3}{c|}{MIIV-2SBMA} & \multicolumn{6}{c|}{Instrument Specific Sargan's $p$}                                     & \multicolumn{6}{c}{Inclusion Probabilities}                       \\
        & Estimate   & SE      & S $p$   & Estimate   & SE     & BMA-S $p$  & $Y_1$ & $Y_2$ & $Y_3$ & $Y_4$ & $Y_7$ & $Y_8$          & $Y_1$ & $Y_2$ & $Y_3$ & $Y_4$ & $Y_7$ & $Y_8$ \\ \hline
Stage 1 & 1.192      & 0.175   & 0.013   & 1.171      & 0.175  & 0.425     & .42 & \textbf{.02}  & .25 & .29 & .49 & \textbf{.02}  & .99  & .17  & 0.15  & .32   & 0.82  & 0.24  \\
Stage 2 & 1.191      & 0.171   & 0.055   & 1.167      & 0.174  & 0.509     & .52 & -    & .30 & .34 & .58 & \textbf{.02}  & .99  & -   & 0.15  & .33   & 0.83  & 0.24  \\
Stage 3 & 1.17      & 0.170   & 0.39   & 1.153      & 0.172  & 0.669       & .67 & -    & .40 & .41 & .71 & -    & .99  & -   & 0.14  & .35   & 0.89  & -     \\ \hline
\end{tabular}%
}
\end{table}

This example differs from the previous one in that both invalid instruments $Y_2$ and $Y_8$ have low inclusion probabilities. Again, it is important to note that low inclusion probabilities reflect instrument weakness conditional on the strength of all other instruments used. Therefore, $Y_2$ and $Y_8$, while having fairly strong bivariate correlations with $Y_5$, are redundant with the other included instruments.

In this case, the Instrument Specific Sargan's Test does not reject valid instruments, but rather indicates that both $Y_2$ and $Y_8$ are invalid when one examines the lowest $p$-value for each stage. This finding corresponds to the results from Simulation 2, where the Instrument Specific Sargan's Test has high specificity to detect weak, invalid instruments. Finally, the BMA Sargan's Test is not significant for any stage of instrument selection, which, as per the simulation study findings, is another indicator that any invalid instruments found are also weak.

\section{Discussion}

In this manuscript we presented the MIIV-2SBMA estimator, an extension of the MIIV-2SLS estimator based on a modification of a procedure first described in \citet{Lenkoski2014}. The development of this estimator was motivated by the three facets of MIIVs for which diagnostics are sorely needed: the presence of invalid MIIVs, determining the specific MIIV that is invalid, and detecting weak MIIVs. Through simulation studies and an empirical example, we demonstrated that the MIIV-2SBMA and associated tests of misspecification and weak instruments provide useful diagnostics to detect these problems.

The MIIV-2SBMA estimator exhibited similar levels of bias as the traditional MIIV-2SLS, with several advantages (see Supplementary Materials). The first is that MIIV-2SBMA is robust to weak instruments. Due to model averaging, models containing weak instruments will be down-weighted, particularly at low sample sizes, as evidenced in both the simulation studies and the empirical example. However, we did find that at low sample sizes ($n=100$), the MIIV-2SBMA estimator had much higher variability, particularly in cases of strong model misspecification. Because of this, we suggest that MIIV-2SBMA is used to perform model misspecification testing, and that the final estimates are compared with standard MIIV-2SLS, which produced less variable estimates.
	
The primary advantage that MIIV-2SBMA has over MIIV-2SLS is the ability to extract Instrument Specific Sargan's Tests and inclusion probabilities. Our Instrument Specific Sargan's Test showed greater power to detect invalid instruments than the traditional Sargan's Test calculated with the complete MIIV set and allows researchers to examine individual instruments for invalidity. Both of our simulation studies suggest that the lowest Instrument Specific Sargan's $p$-value will, with high probability, pinpoint the presence of an invalid instrument, but that the specificity of this approach is greatest when the invalid instrument is also weak. In the case of a strong invalid instrument (e.g., Simulation 1), the specificity of the Instrument Specific Sargan's test was slightly less than in the case of a weak, invalid instrument, but still had relatively good ability to pinpoint an invalid instrument. Given the overall performance of the Instrument Specific Sargan's Test, the use of the smallest $p$-value as the identifier of the true invalid instrument is a promising practice. Additionally, inclusion probabilities give direct assessments of the weakness or strength of a given instrument, making them useful as additional diagnostics, and allows for assessments of conditional weakness, or when instruments are redundant with other instruments. Finally, MIIV-2SBMA is a non-iterative estimator like MIIV-2SLS, and the main gain in computational time is due to evaluating 2SLS solutions over MIIV subsets. For moderately sized MIIV sets, this does not result in a large increase in computation time and the number of subsets evaluated can be changed for extremely large MIIV sets. 

Our simulation study also suggested a use for \citet{Lenkoski2014}'s BMA Sargan's Test. Though this test exhibited lower power than the traditional Sargan's Test, it can be used in tandem with the Instrument Specific Sargan's Tests as an assessment of \textit{meaningful} instrument invalidity. Specifically, a high BMA Sargan's $p$-value in combination with the detection of specific invalid instruments suggests that while there might be misspecification in the model, the misspecification does not result in substantial bias in the given equation when estimated with the MIIV-2SBMA estimator. This reduction in bias is due to the MIIV-2SBMA estimator down-weighting the contribution of weak invalid instruments. Unlike the traditional Sargan's Test, which detects invalid instruments without regard to their potential weakness, the BMA Sargan's Test in combination with the Instrument Specific Sargan's Test provides a diagnostic of potential bias for analysts. More work needs to be done on assessing this particular property.

Our empirical example provides a illustration of how MIIV-2SBMA can be used in practice, by using the Instrument Specific Sargan's Tests to guide model modification.  Furthermore, the example illustrates that MIIV-2SBMA has both a better ability to detect specific invalid instruments, and account for weak invalid instruments, than MIIV-2SLS. 

There are several limitations to the approach presented here. First, the MIIV-2SBMA estimator is similarly impacted by the inclusion of strong, invalid instruments as the MIIV-2SLS making their detection and removal vital for any application. Furthermore, the $g$-prior specification used here is specific to continuously distributed outcomes, and would not be applicable to, for example, binary outcomes. Finally, while the use of the empirical Bayes $g$-prior was chosen for both its good properties and its analytic solutions, different choices for $g$ should be assessed. Specifically, state of the art variable selection priors such as mixture of Zellner-Siow priors should be assessed in future, though this does require more intensive computation \citep{Liang2008}. Finally, while our choice to set the prior mean to the maximum likelihood estimate is consistent with the approach of \citet{Lenkoski2014} and leads to estimates that are close to the maximum likelihood solution, it does not take full advantage of the Bayesian framework used here. If the prior mean of the first stage regression coefficients was, for example, set to 0, the prior would act as a shrinkage estimator, further down-weighting weak instruments. Future research should investigate the specific choice of prior mean.

There are several future directions for the extension of the MIIV-2SBMA estimator. Applying Bayesian model averaging to a generalized methods of moments (GMM) estimator \citep{Hansen1982} would allow for a larger class of models to be fit, and would complement the MIIV-GMM estimator \citep{Bollen2014}. The Instrument Specific Sargan's Test and inclusion probabilities can be used to guide model building, and algorithms should be developed for tracing likely causes of misspecification. Further investigation should be done on the properties of the BMA Sargan's Test, particularly with regard to its utility in assessing for potential bias in parameter estimates. The specificity of the Instrument Specific Sargan's Test should be investigated in a number of other model specifications, as it would be important to determine in which cases does the specificity of the Instrument Specific Sargan's Test suffer.  The sensitivity of the MIIV-2SBMA estimator to other sources of model misspecification should be assessed, as in our simulation studies we restricted our attention to omitted error covariances. Additionally, non-Bayesian model averaging techniques, such as those presented in \citet{sengInstrumentalVariableModel2023}, should be compared with the Bayesian approach detailed here. While both approaches appear to result in unbiased model estimates, it would be useful to evaluate the relative performances of these approaches with respect to sample size and model complexity. Finally, the Instrument Specific Sargan's Test should be compared to modification indices and other classic tests of model misspecification.

In closing, given the additional information provided regarding model misspecification, as well as its robustness to weak instruments, MIIV-2SBMA is an attractive estimator to use for structural equation modeling. It retains all the strengths of the MIIV-2SLS approach, such as bias isolation, computational efficiency and equation level tests of model misspecification, while augmenting them with more specific tests of model misspecification. Further methodological work will extend the MIIV-2SBMA approach to more general non-iterative estimation methods, allowing researcher to evaluate a larger class of models.

\bibliography{References.bib}
\bibliographystyle{apacite}

\end{document}